\documentclass[12pt, letterpaper]{article}
\usepackage[pdftex]{graphicx}
\usepackage{epsfig}
\usepackage{latexsym}
\usepackage{amssymb}
\usepackage{amsmath}
\usepackage{wrapfig}
\usepackage{boxedminipage}
\usepackage{setspace}
\usepackage{subfigure}
\usepackage{topcapt}

\def\N{\cal N}

\DeclareSymbolFont{AMSb}{U}{msb}{m}{n}
\DeclareMathSymbol{\IN}{\mathbin}{AMSb}{"4E}
\DeclareMathSymbol{\IZ}{\mathbin}{AMSb}{"5A}
\DeclareMathSymbol{\IR}{\mathbin}{AMSb}{"52}
\DeclareMathSymbol{\Q}{\mathbin}{AMSb}{"51}
\DeclareMathSymbol{\II}{\mathbin}{AMSb}{"49}
\DeclareMathSymbol{\IC}{\mathbin}{AMSb}{"43}
\DeclareMathSymbol{\IP}{\mathbin}{AMSb}{"50}
\DeclareMathSymbol{\IH}{\mathbin}{AMSb}{"48}
\DeclareMathSymbol\IA{\mathalpha}{AMSb}{"41}
\DeclareMathSymbol\IS{\mathalpha}{AMSb}{"53}

\def\Q{{\cal Q}}

\def\N{{\cal N}}

\begin{document}

\begin{center} 
{\Large \bf  Hot Defect Superconformal Field Theory in an External Magnetic Field }
  \end{center}

\bigskip \bigskip \bigskip

\centerline{\bf Veselin G. Filev}

\bigskip
\bigskip
\bigskip

\centerline{{\it School of Theoretical Physics, Dublin Institute For Advanced Studies}}
\centerline{\it 10 Burlington Road, Dublin 4, Ireland}
\centerline{\small \tt  vfilev@stp.dias.ie}
\bigskip

\bigskip

\bigskip
\bigskip


\begin{abstract} 
\noindent 
In this paper we investigate the influence of an external magnetic field on a flavoured holographic gauge theory dual to the D3/D5 intersection at finite temperature. Our study shows that the external magnetic field has a freezing effect on the confinement/ deconfinement phase transition. We construct the corresponding phase diagram. We investigate some thermodynamic quantities of the theory. A study of the entropy reveals enhanced relative jump of the entropy at the ``chiral" phase transition. A study of the magnetization shows that both the confined and deconfined phases exhibit diamagnetic response. The diamagnetic response in the deconfined phase has a stronger temperature dependence reflecting the temperature dependence of the conductivity. 
We study the meson spectrum of the theory and analyze the stability of the different phases looking at both normal and quasi-normal semi-classical excitations. For the symmetry breaking phase we analyze the corresponding pseudo-Goldstone modes and prove that they satisfy non-relativistic dispersion relation.

\end{abstract}
\newpage \baselineskip=18pt \setcounter{footnote}{0}

\section{Introduction}
Recently a lot of attention has been focused on the properties of flavoured holographic gauge theories and their potential application to realistic phenomenological models. One of the main motivations for such studies is the simple geometric description that holographic models provide of non-perturbative phenomena difficult to study via conventional field theoretical methods. 

Despite the growing spectrum of applications of the  AdS/CFT correspondence \cite{Maldacena:1997re} a major limitation is that realistic field theories do not seem to have simple holographic backgrounds. This is why it is of particular importance to investigate properties of non-abelian gauge theories exhibiting universal behaviour. Particularly interesting is to analyze the phase structure of strongly coupled Yang-Mills theories. An example of such application of the holographic approach is the study of properties of strongly coupled quark-gluon plasmas  (see refs.~\cite{Edelstein:2009,Gubser:2009md} for a recent review). 

The rich and complicated dynamics of strongly coupled non-abelian gauge theories suggests investigation of their response to strong controlling parameters such as temperature, external electromagnetic fields and various chemical potentials. In this regime one may expect that otherwise different theories would exhibit similar behaviour and one can extract valuable results by analyzing holographic gauge theories. This is the context of our present studies. 

Our holographic set up employs holographic gauge theories dual to the Dp/Dq brane intersection. This class of gauge theories has been extensively studied in the literature. The first and most studied example is the D3/D7 intersection \cite{Karch:2002sh}. Early studies of the meson spectrum have been performed in ref.~\cite{Kruczenski:2003be}. The phase structure of the theory at finite temperature has been studied in \cite{Babington:2003vm} and a  confinement / deconfinement phase transition corresponding to topology changing transition of the probe D7-brane has been revealed. More detailed studies of this phase transition have been performed in refs. \cite{Mateos:2006nu}-\cite{Mateos:2007vn}. The phase transition has been classified as a first order one. Universal behaviour of the general Dp/Dq system has been uncovered. Various properties of the general Dp/Dq set up in constant external electric and magnetic fields have been analyzed in refs. \cite{Filev:2007gb}-\cite{Ammon:2009jt}.

One major aspect of the analysis presented in this paper is the phenomena of mass generation in an external magnetic field. This phenomena has been extensively studied in the conventional field theory literature \cite{Gusynin:1994re}-\cite{Hong:1996pv}. The effect was shown to
be model independent and therefore insensitive to the microscopic physics underlying the low energy effective theory. The essence of this effect is the dimensional reduction $D \rightarrow D-2$ ($3+1 \rightarrow 1+1$) in the dynamics of fermion pairing in a magnetic field.

Magnetic catalysis of mass generation has been demonstrated in various 1+2 and 1+3 dimensional field theories. Given the universal nature of this effect it is natural to explore this phenomena in the context of holographic gauge theories. Such studies for the Dp/Dq intersections have been extensively performed for both 1+3 and 1+2 dimensional theories. In ref. \cite{Filev:2009xp} an universal description of the general Dp/Dq system has been attempted. It has been demonstrated that the mechanism of spontaneous symmetry breaking in external magnetic field for the Dp/Dq system is a universal feature of this class of theories.

Particularly interesting are the findings for the D3/D5 set up. Recently this theory received a great deal of attention and emphasis has been made of the potential application of this brane configuration in describing qualitative properties of $1+2$
dimensional condensed matter systems (see for example refs.~\cite{{O'Bannon:2008bz},{Myers:2008me},{Evans:2008nf}}). In ref. \cite{Filev:2009xp} it has been shown that the pseudo goldstone modes of this theory satisfy non-relativistic dispersion relation. The low energy effective action has been obtained. To the computed order, it agrees to the effective action of spin waves in a ferromagnetic. The existence of a single time derivative term in the effective action confirms the potential phenomenological applications of this model.

In this paper we extend our previous investigation of the D3/D5 set up in an external magnetic field to the case of finite temperature. However, we are not only interested in expanding the description of the pseudo-goldstone modes to the finite temperature case but also to study the effect that the external magnetic field has on the confinement/deconfinement phase transition. This is why we are interested in both the small and large bare mass sectors of the theory. Similar studies of the D3/D7 set up have been performed in \cite{Erdmenger:2007bn,Albash:2007bk}. It has been shown that the magnetic field has a freezing effect on the phase transition and for sufficiently strong magnetic field the deconfined phase cease existence (for any bare mass). Our studies reveal the same qualitative behaviour in the case of the D3/D5 set up. However there are some differences that are of potential interest.

Let us summarize the content of the paper:

Section 2  of this work is separated into two subsections. In the first subsection we provide a brief description of the holographic set up. The parameterization of the background geometry that we use, the basic extracts from the AdS/CFT dictionary needed for the holographic description as well as the method of introducing external magnetic field, are presented. The second subsection studies the equation of state of the system in the condensate versus bare mass plane. The study involves both analytic and numerical techniques. The analytic results are for large bare masses. An asymptotic expression for the condensate as a function of the magnetic field, the temperature and the bare mass is obtained. It is demonstrated that unlike the 1+3 dimensional case (the D3/D7 set up) in the 1+2 case the competing effect of the magnetic field and the temperature is present even at large bare masses. This gives the possibility to tune the condensate of the theory to zero for arbitrarily large masses. The reason for that could be that unlike in the D3/D7 set up the external magnetic field does not require new counter terms in the holographic regularization of the D-brane action. For small bare masses we perform a numerical analysis. The freezing effect of the magnetic field is demonstrated. The existence of critical ratio of the temperature and the square root of the magnetic field beyond which the phase transition disappears is demonstrated.

Section 3 of this parer is dedicated to the thermodynamic analysis of the theory.
A holographic renormalization of the D5-brane action in the spirit of \cite{Karch:2005ms} is performed and the free energy of the theory is obtained. The proper thermodynamic ensemble is identified and the phase diagram of the theory in various coordinates is constructed. The temperature dependence of the condensate at zero bare mass is explored as an order parameter for the ``chiral" phase transition. A study of the entropy is performed. It is demonstrated that at zero bare mass the jump of the entropy is enhanced relative to the jump of the entropy at large bare masses. A possible explanation is that at zero bare mass the confinement/deconfinement phase transition is also a chiral phase transition. Finally, the magnetization of the theory is computed. It is demonstrated that both the confined and deconfined phases are diamagnetic. In the confined phase the diamagnetic response of the theory is almost independent on the temperature. At the phase transition the magnetization has a large negative jump and a very strong diamagnetic response possibly because the theory is in a conductive phase. Interestingly the diamagnetic response of the deconfined phase depends strongly on the temperature and vanishes inversely with the temperature. The strong temperature dependence reflects the temperature dependence of the conductivity. At large temperature the susceptibility is deduced by dimensional analysis reflecting the return to conformality. 

Section 4 of this work studies the meson spectrum of the theory. 
First the spectrum of fluctuations corresponding to the radial coordinate in the transverse $\IR^3$ are considered and the stability of the theory is analyzed. Both normal and quasi-normal excitations are studied. It is demonstrated that the thermodynamically stable phases are tachyon free. It is also verified that phases with negative heat capacity have tachyonic spectrum and are thus unstable. The existence of metastable phases accessible by supercooling is inferred. At zero bare mass the case of zero magnetic field and finite temperature is analyzed along the lines of ref. \cite{Hoyos:2006gb}. The corresponding Heun equation is solved employing the method of continued fractions. Next, the spectrum corresponding to fluctuations along the angular coordinates of the transverse $\IR^3$ is analyzed. Only the confined phase is considered. The qualitative features of the spectrum are the same as for the zero temperature case studied in \cite{Filev:2009xp}. At large bare mass a Zeeman splitting of the energy levels is observed. For small bare mass the pseduo-Golstone mode is analyzed. A non-relativistic dispersion relation is observed. The analysis of the pseudo-Goldstine modes is performed both numerically and analytically. 

We end the paper with a brief conclusion in Section 5.

\section{Preliminaries}
\subsection{General set up}
 Let us consider the supergravity background corresponding to the near horizon limit of $N_c$ coincident D3-branes at finite temperature: 
\begin{eqnarray}
ds^2/\alpha'&=&-\frac{(4r^4-b^4)^2}{4R^2r^2(4r^4+b^4)}dt^2+\frac{4r^4+b^4}{4R^2r^2}d{\vec x}^2+\frac{R^2}{r^2}(d\rho^2+\rho^2d\Omega_2^2+dl^2+l^2d\tilde\Omega_{2}^2);\label{background}\\
e^{\Phi}&=&\frac{1}{g_s};~~~C^{(4)}_{01234}=\frac{1}{g_s}\frac{(4r^4+b^4)^2}{16R^4r^4};\ ,\nonumber \\
\rm{where}~~d\Omega_2&=&d\alpha^2+\cos\alpha^2d\beta^2;~~~d\tilde\Omega_{2}^2=d\psi^2+\cos^2\psi d\phi^2;~~~r^2=\rho^2+l^2;\ .\nonumber
\end{eqnarray}
The metric in equation (\ref{background}) corresponds to the AdS$_5\times S^5$ black hole in a Poincare patch. Its extremal limit  is holographically dual to an ${\cal N}=4$ $SU(N_c)$ Supersymetric Yang-Mills theory in $1+3$ dimensions \cite{Maldacena:1997re}. The radius of the asymptotically AdS$_5$ space is related to the t'Hooft coupling of the gauge theory via $R^2=2\lambda\alpha'$. The Hawking temperature of the AdS$_5$ black hole is given by $T=b/\pi R^2$ and is interpreted as the temperature of the holographically dual gauge theory. For the choice of radial coordinate considered in equation (\ref{background}) the subspace transverse to the volume of the background colour branes is conformally $\IR^6$. We have further split the transverse $\IR^6$ subspace to the product $\IR^3\times\IR^3$ and introduced spherical coordinates $\rho,\Omega_2$ and $l,\tilde\Omega_2$ in the corresponding subspaces.

Next we introduce $N_f$ probe D5-branes extended along the $t,x_1,x_2,\rho,\Omega_2$ directions of the geometry. In the extremal limit such embeddings preserve half of the original supersymmetries of the background. The lowest energy sector of strings stretched between the D3 and D5 branes gives rise to  $N_f$ fundamental $\N=2$ hypermultiplets
confined on a $1+2$ dimensional defect \cite{Karch:2000gx,DeWolfe:2001pq,Erdmenger:2002ex}. The asymptotic separation of
the D3 and D5 --branes in the transverse $\IR^3$ subspace is parameterized
by $l$ and is proptional to the bare mass of the fundamental $\N=2$ hypermultiplets.
If we consider the following anzatz for the D5--brane emebedings:
\begin{equation}
l=l(\rho)\ ;~~~\psi=0\ ;~~~\phi=0\ ,\label{anz2}
\end{equation}
the exact relation is given by:
\begin{equation}
m\equiv l(\infty)=(2\pi\alpha')m_q\ ,
\end{equation}
where $m_q$ is the bare mass of the fundamental field in the corresponding gauge theory. If the D3 and D5
branes overlap, the corresponding fundamental fields are massless and the theory has a global $SO(3)_R\times SO(3)$
symmetry corresponding to independent rotations in the two $\IR^3$ subspaces the transverse $\IR^6$ space. A non-trivial profile of the D5--brane $(l(r)\neq0)$ in the
transverse $\IR^3$ subspace would break the global symmetry down to
$SO(3)_R\times U(1)$, where $U(1)$ is the little group in the transverse
$\IR^3$ subspace. Furthermore, if the D5-brane wraps a shrinking $S^2$ cycle and closes at some radial (coordinate) distance above the horizon this distance is interpreted as the dynamically generated mass of the theory \cite{Mateos:2007vn}.

However the gauge theory dual to the D3/D5 system is conformal at zero bare mass and does not exhibit spontaneous symmetry breaking. One way to catalyze spontaneous symmetry breaking is to introduce a constant magnetic field to the theory. The effect of mass generation in external magnetic field has been widely studied on field theory side refs.~\cite{Gusynin:1994re}-\cite{Hong:1996pv}. In the context of AdS/CFT the effect of magnetic catalysis in holographic gauge theories has been studied in \cite{Filev:2007gb}-\cite{Johnson:2009ev}. In order to introduce magnetic field to the holographic set up we turn on a constant $B$-field along the $x_1,x_2$ directions of the geometry:
\begin{equation}
B_{(2)}=H dx_1\wedge dx_2\ .
\end{equation}
On field theory side this corresponds to introducing constant magnetic field $H/(2\pi\alpha')$ perpendicular to the plane of the defect. The D5--brane embedding is determined by the DBI
action:
\begin{eqnarray}
S_{\rm{DBI}}=-N_f\mu_{5}\int\limits_{{\cal M}_{6}}d^{6}\xi e^{-\Phi}[-{\rm det}(G_{ab}+B_{ab}+2\pi\alpha' F_{ab})]^{1/2}\  . \label{DBI2}
\end{eqnarray}
Where $G_{ab}$ and $B_{ab}$ are the pull-back of the metric and the $B$-field respectively and $F_{ab}$ is the gauge field on the D5--brane.

With the anzatz (\ref{anz2}) the Lagrangian is given by:
\begin{equation}
{\cal L}\propto \rho^2\left(1-\frac{b^4}{4r^4}\right)\left(1+\frac{b^4}{4r^4}\right)^{1/2}\left(1+\frac{16R^4H^2r^4}{(4r^4+b^4)^2}\right)^{1/2}\sqrt{1+l'^2}\cos\alpha\ .\label{lagr2}
\end{equation}
For large $\rho\gg b$ the Lagrangian (\ref{lagr2}) asymptotes to:
\begin{equation}
{\cal L}\propto\rho^2\sqrt{1+l'^2}\cos\alpha\ ,
\end{equation}
which has a general solution of the form:
\begin{equation}
l(\rho)=m+\frac{c}{\rho}+\dots\label{dict}
\end{equation}
as we mentioned above the parameter $m$ is the asymptotic separation of the D3 and D5 branes and corresponds to the bare mass of the dual gauge theory. The AdS/CFT dictionary suggests \cite{Karch:2002sh} that the parameter $c$ is proportional to the condensate of the fundamental fields $\langle\bar\psi\psi\rangle$. The exact relation is given by \cite{Mateos:2006nu}:
\begin{equation}
\langle\bar\psi\psi\rangle=-8\pi^2\alpha'\frac{\mu_5}{g_s}c\ . \label{conddef}
\end{equation}
By solving the equations of motion of the probe D5-brane and extracting the asymptotic behaviour at infinity one can generate the equation of state of theory in the condensate versus bare mass plane. This is one of the main tools that the holographic set up is providing and will be one of the main technique that we use to construct the phase diagram of the theory. 

\subsection{The condensate of the theory}
In this section we study the fundamental condensate of the theory. We focus on the dependence of the condensate as a function of the bare mass for fixed temperature and magnetic field. We explore the effect of the magnetic field on the confinement/deconfinement phase transition of the fundamental matter. We show that the magnetic field has a freezing effect on the transition. We also show that for sufficiently strong magnetic field the theory develops a negative condensate at zero bare mass and hence the global $SO(3)$ symmetry is spontaneously broken. We will first focus on the regime of large bare masses $m\gg b, R\sqrt{H}$, which can be treated analytically. 

\subsubsection{Exact results for large bare masses.}

As we discussed above large bare mass correspond to large separation between the color and flavour branes. In this regime the probes are nearly BPS objects, their embeddings are very close to the trivial $l\equiv 0$ one and we can expand:
\begin{equation}
l(\rho)=a+\xi(\rho)\ ,
\label{eqemb}
\end{equation}
where $\xi(\rho)\ll m$ and $a\equiv l(0)$ so that we have:
\begin{equation}
 \xi(0)=\xi'(0)=0 \ .
 \label{bndr}
 \end{equation}
 On the other side the radial coordinate $r\gg b$ remains large along the D5-brane embedding hence we can expand the lagrangian (\ref{lagr2}) in powers of $b/r$. To leading order we obtain the following equation of motion:
\begin{equation}
\partial_{\rho}(\rho^2\xi'(\rho))=a\rho^2\left(\frac{b^4-4H^2R^4}{2(\rho^2+a^2)^3}+\frac{5b^8+40b^4H^2R^4+16H^4R^8}{8(\rho^2+a^2)^5}\right)+\dots
\label{eqnlin}
\end{equation}
After solving for $\xi(\rho)$, imposing the boundary conditions (\ref{bndr}) and substituting in (\ref{eqnlin}) we can extract the following expressions for the bare mass $m$ and fundamental condensate $c$:
\begin{eqnarray}
m&=&a+\frac{b^4-4H^2R^4}{8a^3}+O\left(\frac{1}{a^7}\right);\label{largemcond}\\
c&=&-\frac{\pi(b^4-4H^2R^4)}{32a^2}-\frac{5\pi(5b^8+40b^4H^2R^4+16H^4R^8)}{4096a^6}+O\left(\frac{1}{a^{11}}\right); \ .\nonumber
\end{eqnarray}
The first equation in (\ref{largemcond}) can be easily inverted to obtain our final expression for the fundamental condensate as a function of the bare mass valid for large $m$:
\begin{equation}
\langle\bar\psi\psi\rangle\propto -c=\frac{\pi(b^4-4H^2R^4)}{32m^2}+\frac{\pi(57b^8-56b^4H^2R^4+592H^4R^8)}{4096m^6}+O\left(\frac{1}{m^{11}}\right);
\end{equation}
A few comments are in order:

 It is interesting that even for large bare masses the competition between the effect of magnetic field and finite temperature is still present and the condensate can be either negative or positive depending on the ratio $b/R\sqrt{H}$. This is to be compared to the $1+3$ dimensional case, studied in \cite{Erdmenger:2007bn,Albash:2007bk} (using the D3/D7 set up),  where the leading contribution to the condensate is negative and proportional to $R^4H^2/m$. As we are going to see later this seems to be related to the fact that in the $1+2$ dimensional case the external magnetic field does not lead to new divergences in the on-shell action, while in the $1+3$ dimensional case there is a logarithmic divergence due to the magnetic field \cite{Albash:2007bk}. 

To extend our analysis to the regime of large temperature and magnetic field we need to employ numerical techniques.
 
 \subsection{Numerical results}
In this section we solve numerically the equations of motion for the D5-brane embedding derived from (\ref{lagr2}) and explore the dependence of the condensate on the bare mass for fixed ratio of the external magnetic field and temperature of the system. We show that qualitatively the behavior of the system is the same as in $1+3$ dimensions for the D3/D7 setup \cite{Erdmenger:2007bn,Albash:2007bk}. 

Note that the background geometry (\ref{background}) has a horizon at $r_0^2=b^2/2$ and the probe D5-brane embeddings split into two classes:  The first class are Minkowski embeddings which wrap a shrinking $S^2$ in the $S^5$ part of the geometry and close at some finite distance above the balck hole. The second class of embeddings are Black hole type of embeddings that fall into the black hole and have an induced horizon on their worldvolume. The transition from one class of embeddings to the other class of embeddings is a topology change transition that corresponds to a first order confinement/deconfinement phase transition for the fundamental matter \cite{Babington:2003vm,Mateos:2006nu,Albash:2006ew}. There is also a critical embedding separating the two classes of embeddings that has a conical singularity at the horizon of the geometry. A more detailed analysis of the dual gauge theory near the state corresponding to the critical embedding reveals a discrete self-similar behaviour of the theory. In this regime the theory is characterized by the existence of multiple (thermodynamically unstable) phases manifested by a double logarithmic structure of the equation of state curve \cite{Mateos:2006nu} in the condensate versus bare mass plane. We shall not analyze this unstable regime of the theory any further. Instead we shall focus on the influence of the external magnetic field on the confinement/deconfinement phase transition. As we are going to show the effect of the magnetic field is to decrease the critical temperature at which this transition takes place. Furthermore we shall demonstrate that for sufficiently strong magnetic field the phase transition cease existence and Minkowski embeddings are the only thermodynamically stable ones. 

\begin{figure}[h] 
   \centering
   \includegraphics[width=6.5cm]{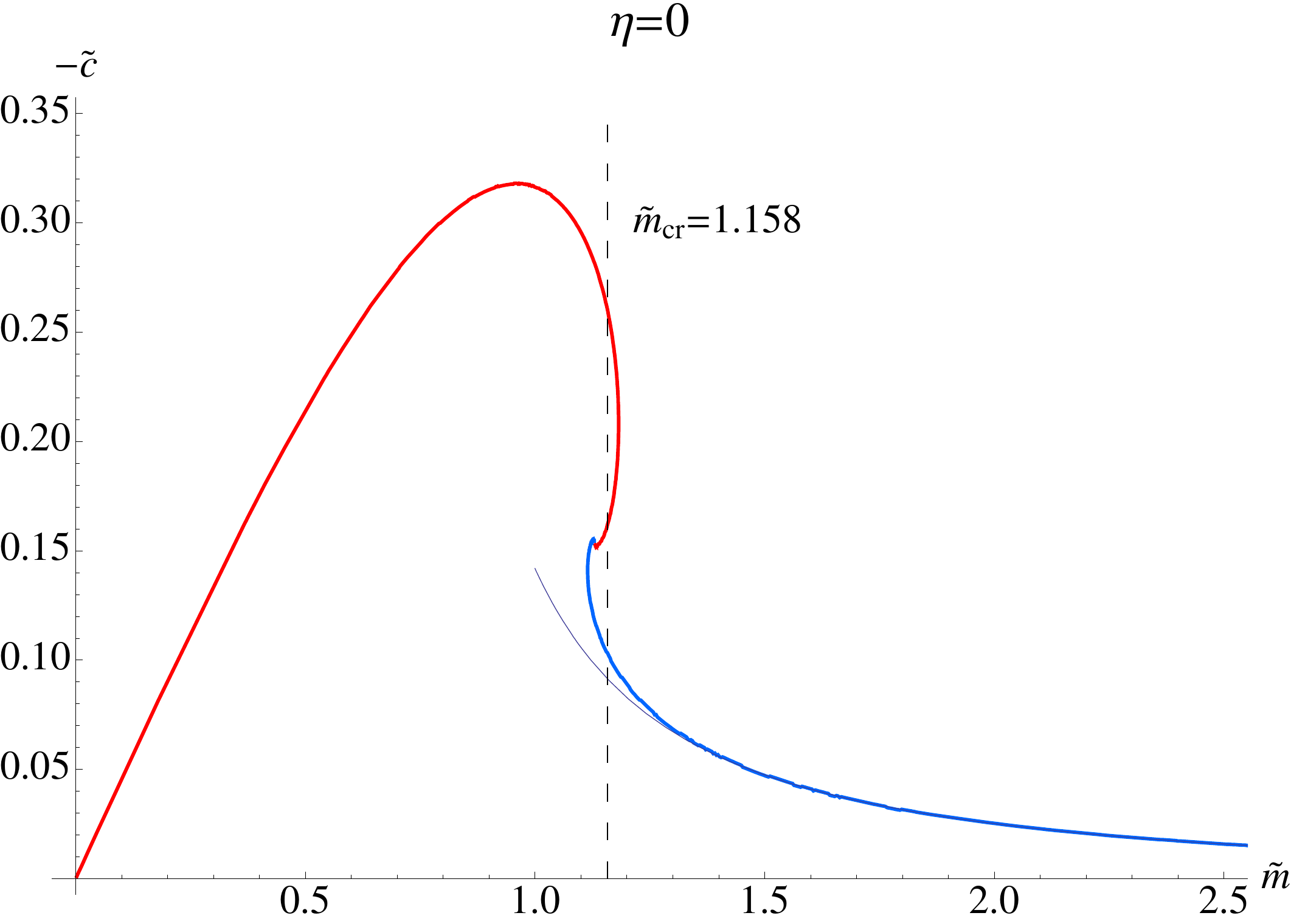}
      \hspace{1cm}
    \includegraphics[width=6.5cm]{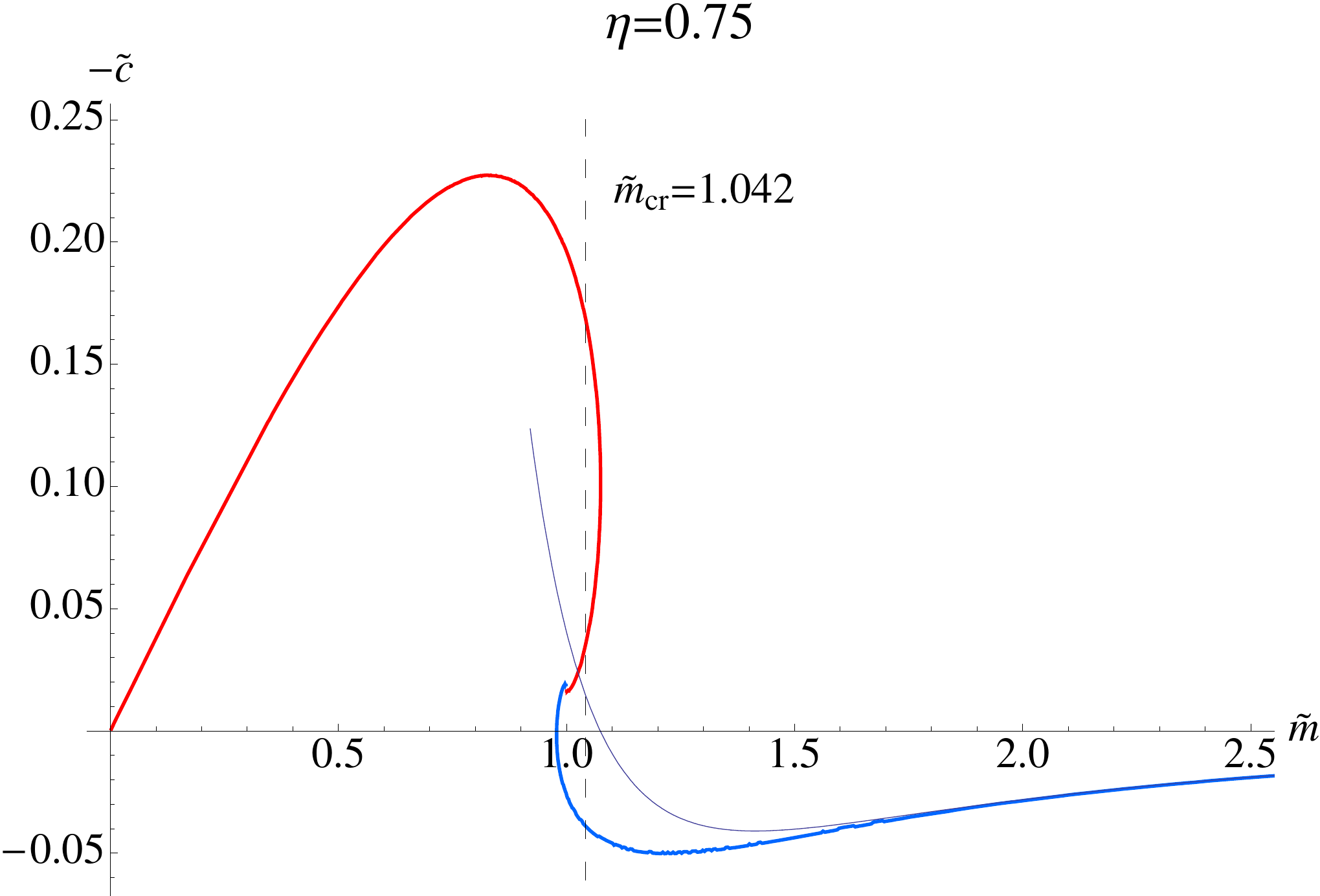}
      \includegraphics[width=6.5cm]{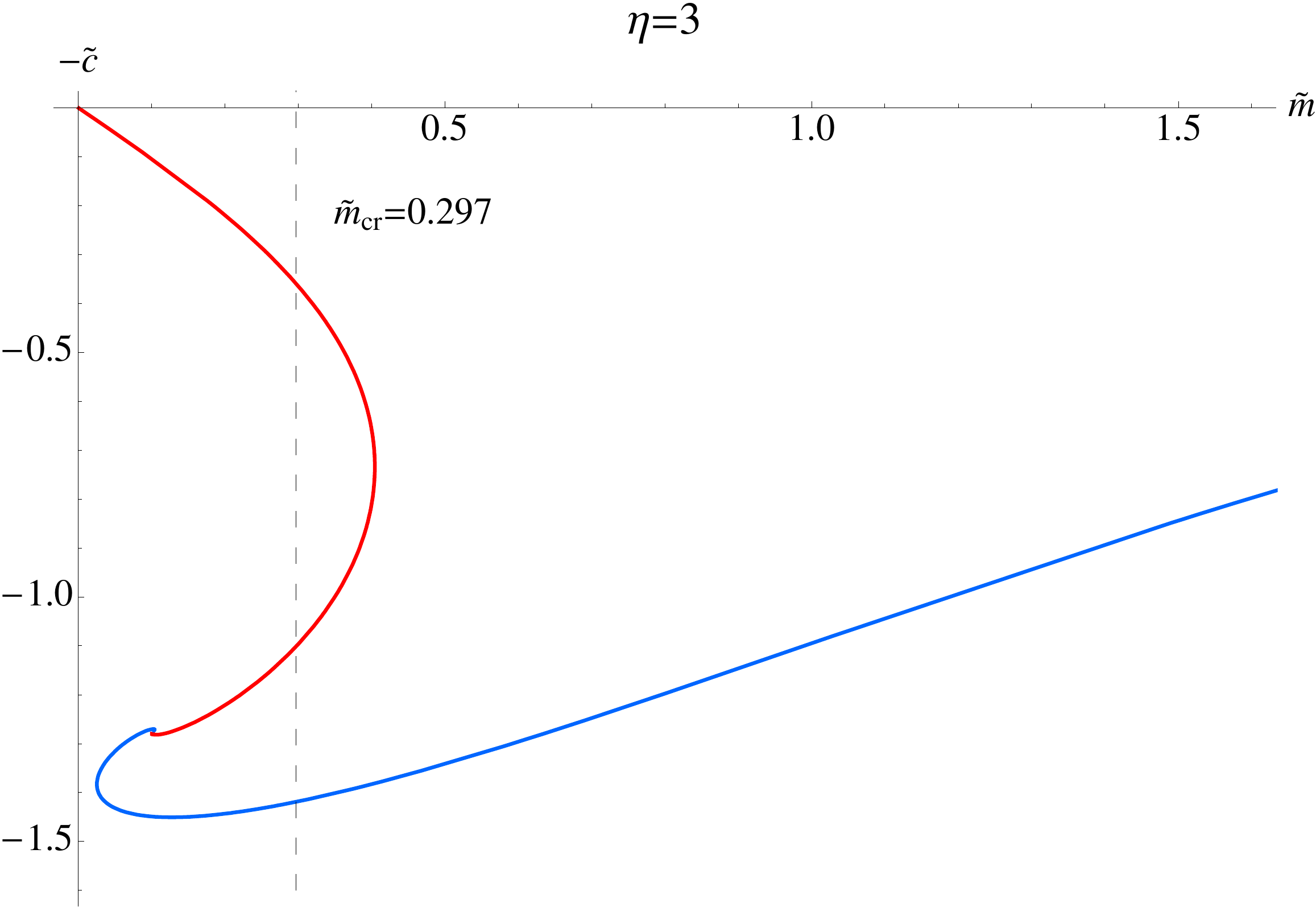}
         \hspace{1cm}
        \includegraphics[width=6.5cm]{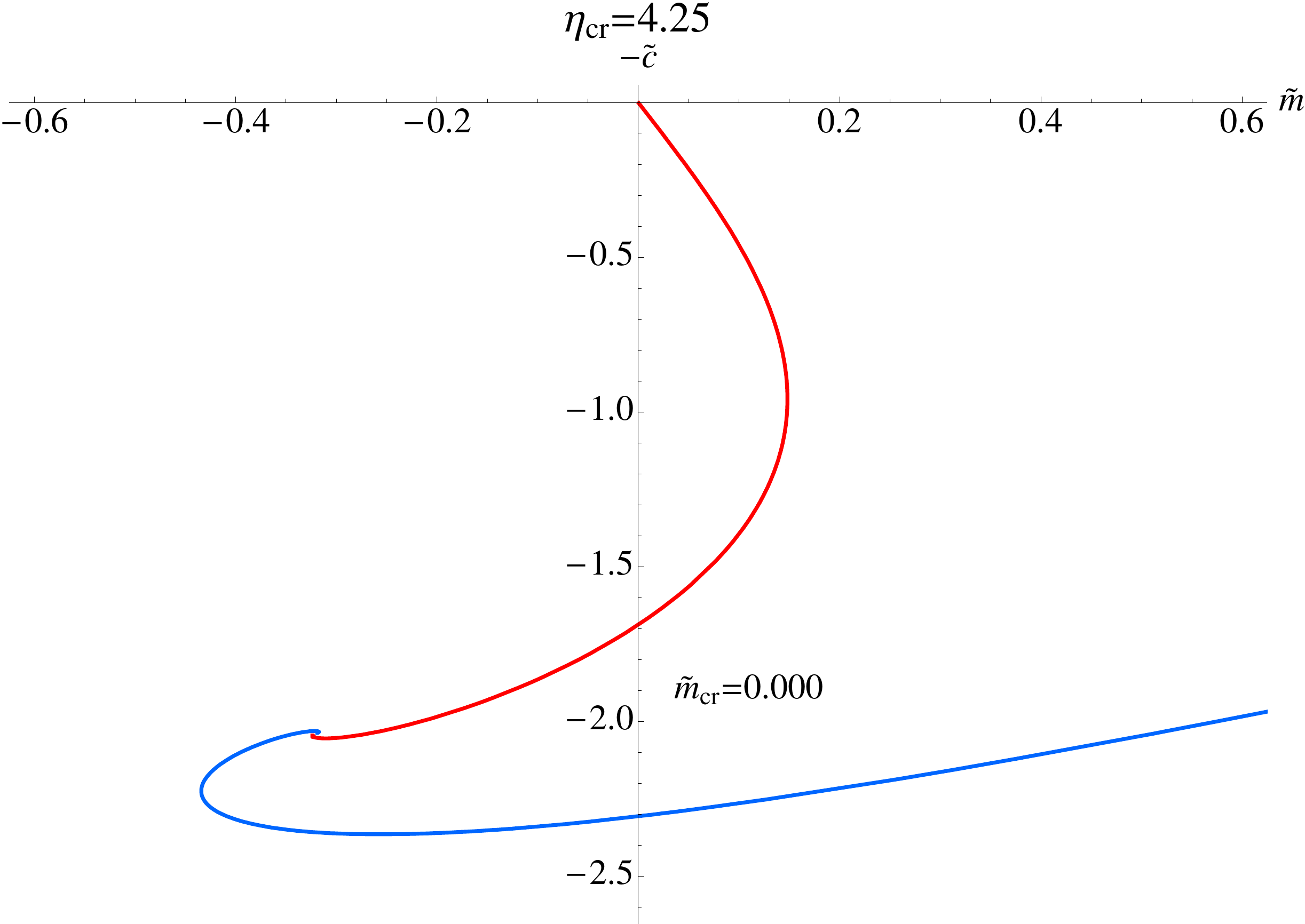}
   \caption{\small As one can see the critical bare mass decreases as the magnetic field increases and for $\eta_{cr}\approx4.25$ it vanishes. Beyond this point the lowest positive branch is the stable one and at $\tilde m=0$ there is a negative condensate breaking the global $SO(3)$ symmetry.}
   \label{fig:fig1}
\end{figure}

For our numerical analysis it is convenient to define the following dimensionless variables:
\begin{eqnarray}
\tilde\rho={\rho}/{m};~~~\eta=HR^2/b^2;~~~\tilde m=m/b;~~~\tilde c=c/b^2;~~~\tilde l(\tilde\rho)=\frac{1}{b}l(\rho/b)=\tilde m+\tilde c/\tilde\rho+\dots\label{bezrazm}
\end{eqnarray}
Upon substitution into the Lagrangian (\ref{lagr2}) we obtain:
\begin{equation}
{\cal \tilde L} \propto \tilde\rho^2\left(1-\frac{1}{4\tilde r^4}\right)\left(1+\frac{1}{4\tilde r^4}\right)^{1/2}\left(1+\frac{16\eta^2\tilde r^4}{(4\tilde r^4+1)^2}\right)^{1/2}\sqrt{1+\tilde l'^2}\cos\alpha\label{dimleslag}
\end{equation}
To solve numerically the equation of motion obtained from (\ref{dimleslag}) we impose the following boundary conditions:
\begin{eqnarray}
&&\tilde l(0)=\tilde l_0;~~~\tilde l'(0)=0;~~~\text{for Minkowski embeddings,}\\
&&\tilde l(\tilde\rho)|_{\rm{e.h.}}=\tilde l_0;~~~\tilde l'(\tilde\rho)|_{\rm{e.h.}}=\frac{\tilde l_0}{\tilde\rho}\Big|_{\rm{e.h.}};~~\text{for Black Hole embeddings.}\nonumber
\end{eqnarray}
Using equation (\ref{dict}) we can extract the condensate and bare mass parameters $\tilde m,\tilde c$. The resulting plots for $\eta=0, 0.75, 2, 4.25$ are presented in figure \ref{fig:fig1}. One can clearly see a first order phase transition pattern. 
It is also evident that the critical bare mass $\tilde m_{cr}$ decreases as the magnetic field increases and for $\eta_{cr}\approx 4.25$ it vanishes. Beyond this point (for $\eta>\eta_{cr}$) the lowest positive branch of the condensate vs. bare mass plot is the stable one. Furthermore at $\tilde m=0$ the theory develops a non-negative condensate and hence the global $SO(3)$ symmetry is spontaneously broken. It turns out that this symmetry breaking phase exists even for smaller values of the magnetic field ($3<\eta<4.25$). The analysis of the meson spectrum in Section 4 will show that this is in fact a metastable phase of the theory that becomes stable at $\eta=\eta_{cr}$. Note also that there are more than one phases with broken global symmetry and zero bare mass. However analysis of the free energy shows (see Section 3) that the phase corresponding to the lowest negative value of the fundamental condensate is the stable phase. 

For large ratios of the external magnetic field and temperature of the system ($\eta>\eta_{\rm{cr}}$) the deconfined phase is parametrically suppressed and the qualitative behaviour of the theory approaches the one studied in ref. \cite{Filev:2009xp}. In this limit the condensate curve has a spiral structure and the stable phase of the theory corresponds to the lowest positive branch of the spiral.

The critical mass $\tilde m_{cr}$ in figure \ref{fig:fig1} was determined by evaluating the free energy of the system (alternatively we could use the {\it equal areas law} \cite{Albash:2007bk}), which suggests renormalizing the euclidean on-shell action of the probe brane and analyzing the thermodynamic properties of the theory. Analyzing these properties and constructing the phase diagram of the theory will be the subject of the next section. 

\section{Thermodynamic analysis}
Let us begin by calculating the free energy of the theory.
\subsection{Free energy}
Since our theory is at fixed temperature $T$ and magnetic field $B$ the density of of the thermodynamic potential describing the ensemble satisfies \cite{Albash:2007bk}:
\begin{equation}
dF=-SdT-\mu dB\ .
 \end{equation}
Here $S$ is the entropy density of the system, $B=H/(2\pi\alpha')$ is the magnetic field and $\mu$ is the magnetization.  

\begin{figure}[h] 
   \centering
   \includegraphics[width=6.5cm]{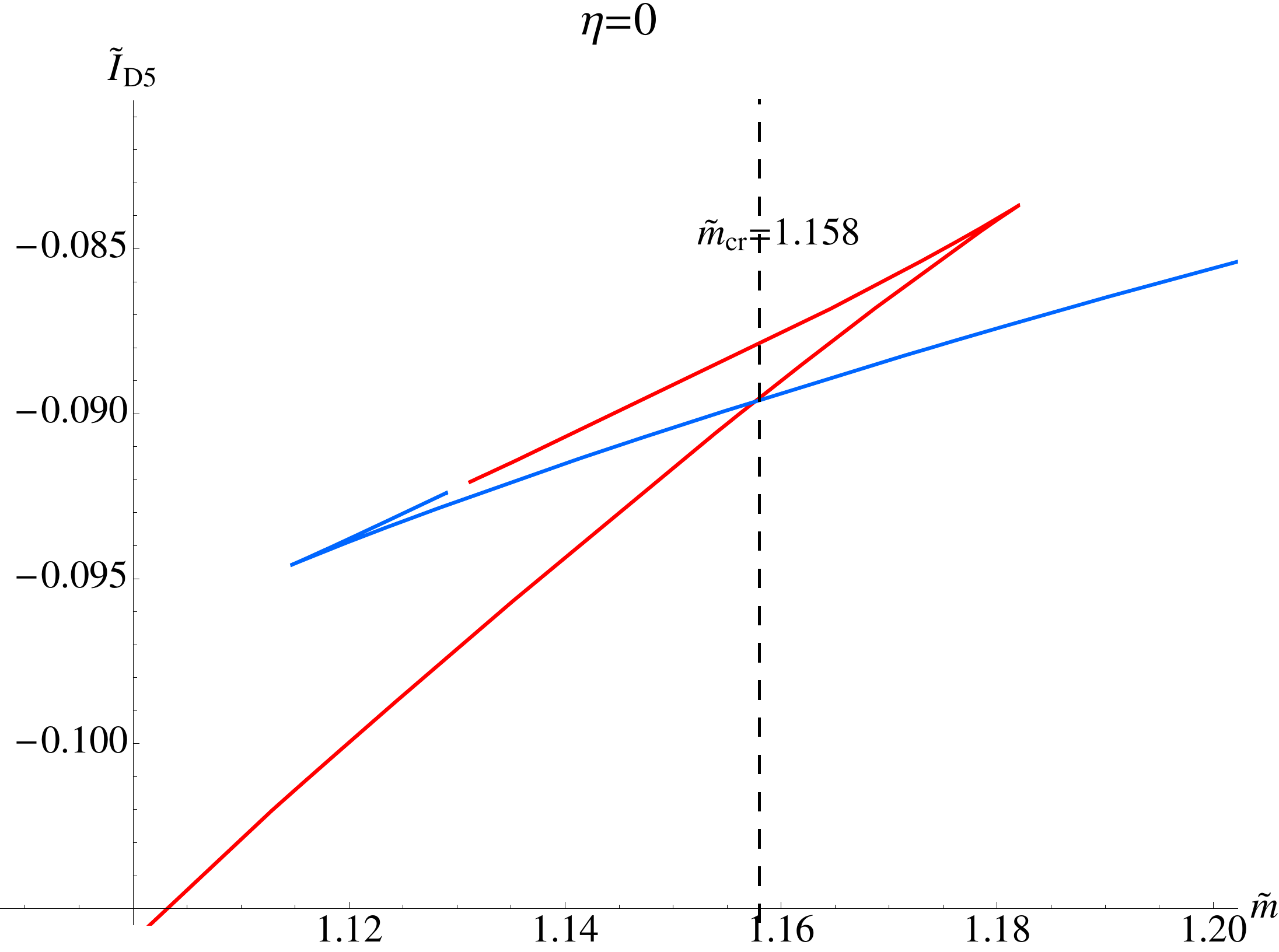}
   \hspace{1.5cm}
    \includegraphics[width=6.5cm]{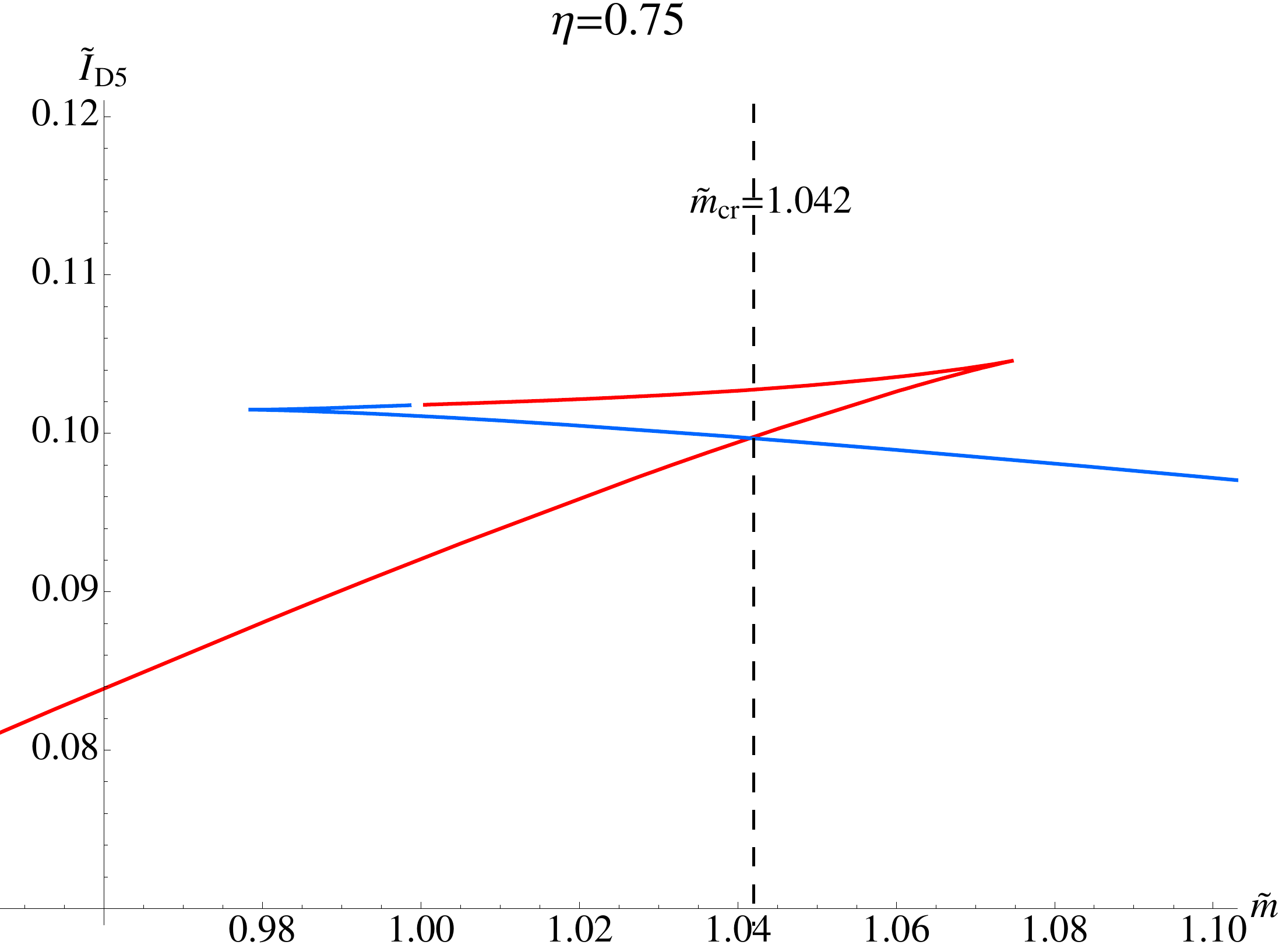}
      \includegraphics[width=6.5cm]{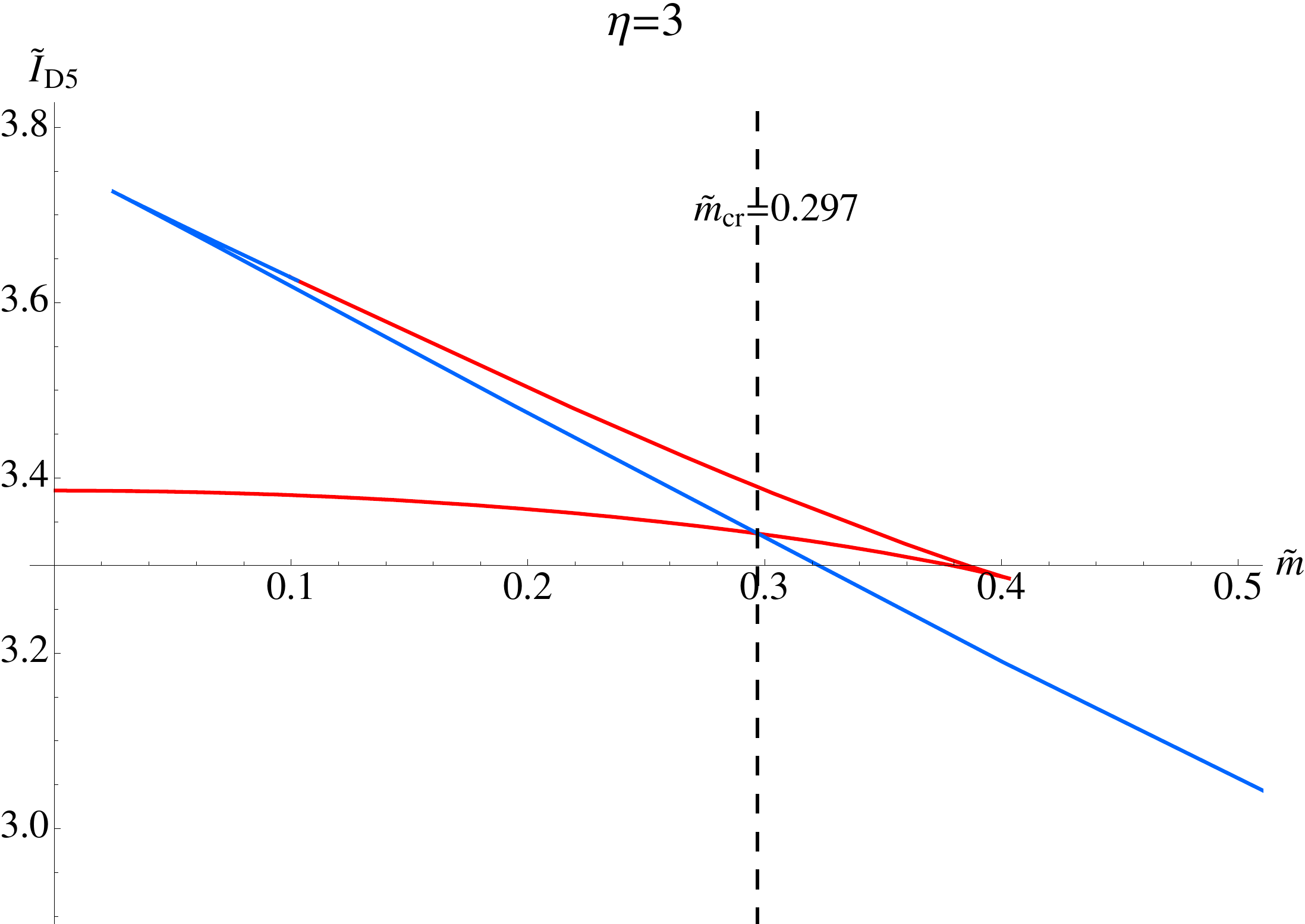}
         \hspace{1.5cm}
        \includegraphics[width=6.5cm]{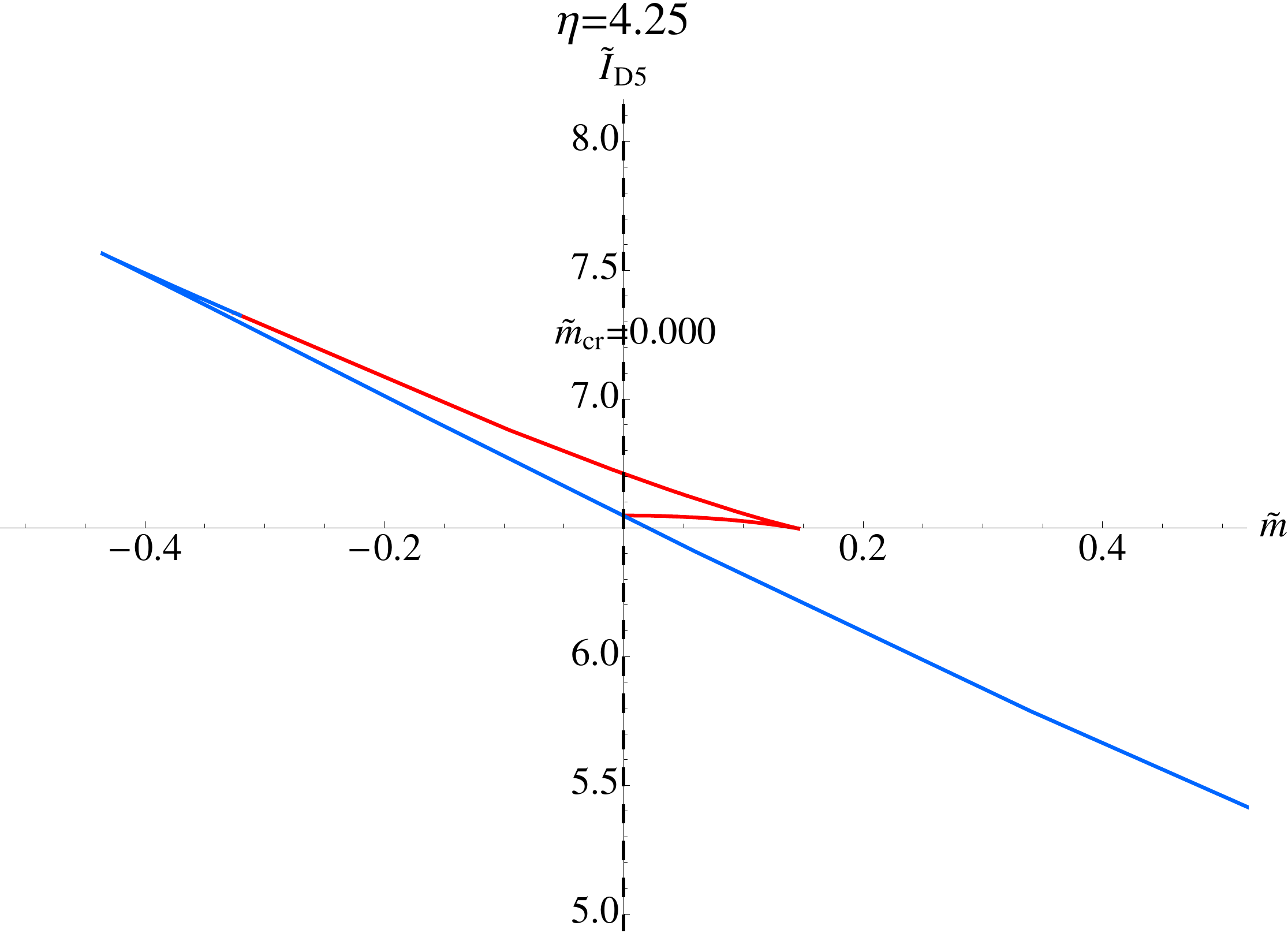}
   \caption{\small Plots of $\tilde I_{\rm{D5}}$ vs. $\tilde m$ for $\eta=0,.75,3,4.25$. The critical value $\tilde m_{\rm{cr}}$ is determined by the position of the kink of $\tilde I_{\rm{D5}}$.}
   \label{fig:fig2}
\end{figure}

Following ref.~\cite{Mateos:2007vn} we will identify the regularized
wick rotated on-shell lagrangian of the D5--brane with the free energy of
the theory. Let us introduce a cut-off at infinity, $\rho_{\max}$, The
wick rotated on-shell lagrangian is given by:
\begin{equation}
I_{\rm{D5}}=N_f\frac{\mu_5}{g_s}4\pi V_3b^3\int\limits_{\tilde\rho_{\rm{min}}}^{\tilde \rho_{\rm{max}}} d\tilde \rho\tilde \rho^2\left(1-\frac{1}{4\tilde r^4}\right)\left(1+\frac{1}{4\tilde r^4}\right)^{1/2}\left(1+\frac{16\eta^2\tilde r^4}{(4\tilde r^4+1)^2}\right)^{1/2}\sqrt{1+\tilde l'^2}\ ,
\label{actionwick}
\end{equation}

where $V_3=\int d^3x$ and $\tilde l(\tilde \rho)$ is the solution of the equation of motion derived from (\ref{dimleslag}).  The parameter $\tilde \rho_{\rm{min}}=0 $ for Minkowski embeddings and $\tilde\rho_{\rm{min}}=\tilde\rho|_{\rm{e.h.}}$ for Black hole embeddings. It is easy to check that for large $\rho_{max}$ we have:
\begin{equation}
b^3\int\limits^{\tilde \rho_{\rm{max}}} d\tilde \rho\tilde \rho^2\left(1-\frac{1}{4\tilde r^4}\right)\left(1+\frac{1}{4\tilde r^4}\right)^{1/2}\left(1+\frac{16\eta^2\tilde r^4}{(4\tilde r^4+1)^2}\right)^{1/2}\sqrt{1+\tilde l'^2}=\frac{1}{3}\rho_{\rm{max}}^3+O\left(\frac{1}{\rho_{\rm{max}}}\right)\ .
\end{equation}
Note that the the constant $B$-field does not introduce new divergencies to the on-shell action. This is different from the $1+3$ dimensional case studied via the D3/D7 set up, where a logarithmic divergency is introduced proportional to $R^4H^2$. The $\rho_{\rm{max}}^3$ divergency is present even at zero temperature and magnetic field and can be removed by the addition of appropriate boundary terms. These terms have been computed in \cite{Karch:2005ms}, where an elegant renormalization procedure for general Dp/Dq intersections was developed. One can easily  check that in our coordinates the counter terms action boils down to $-\frac{1}{3}\rho_{\rm{max}}^3$. It is then convenient to define the following finite dimensionless quantity:
\begin{equation}
\tilde I_{\rm{D5}}=\int\limits_{\tilde \rho_{\rm{min}}}^{\infty} d\tilde \rho\tilde \rho^2\left\{\left(1-\frac{1}{4\tilde r^4}\right)\left(1+\frac{1}{4\tilde r^4}\right)^{1/2}\left(1+\frac{16\eta^2\tilde r^4}{(4\tilde r^4+1)^2}\right)^{1/2}\sqrt{1+\tilde l'^2}-1\right\}-\frac{1}{3}\tilde\rho_{\rm{min}}^3\ .\label{D5dimles}
\end{equation}
Note that $\tilde I_{\rm{D5}}$ implicitly depends on $\tilde m$ trough the on-shell form of $\tilde l(\tilde\rho)$. Our final expression for the density of the free energy is then \footnote{Note that to avoid repetition we will sometimes refer to the densities of the thermodynamic quantities of interest by omitting the term density}:
\begin{equation}
F=N_f\frac{\mu_5}{g_s}4\pi b^3\tilde I_{\rm{D5}}(\tilde m,\eta^2)\ . \label{Free-energy}
\end{equation}
Using equation (\ref{Free-energy}) we can calculate the free energy density of the dual gauge theory in the functional form $F(T,B,m_q)=T^3f(\frac{m_q}{T},\frac{B}{T^2})$, therefore the quantity $\tilde I_{\rm{D5}}\propto F/T^3$ can be used to determine the preferred phase at a given temperature. It is straightforward to calculate the critical bare mass parameter $\tilde m_{\rm{cr}}$ at which the confinement/deconfinement phase transition takes place. Plots of $\tilde I_{\rm{D5}}$ vs. $\tilde m$ for $\eta=0,0.75,3,4.25$ (the values from figure \ref{fig:fig1}) are presented in figure \ref{fig:fig2}. The critical values $\tilde m_{\rm{cr}}$ are determined by the kinks of the free energy. If we refine our numerical analysis and scan through all possible values of $\eta$ for witch there is a phase transition, namely $0\le\eta\leq 4.25$, we can generate the phase diagram of the theory.

\subsection{Phase diagram}
In order to construct the phase diagram of the theory we generate numerically the critical curve $\tilde m_{\rm{cr}}(\eta)$ in the $\tilde\eta$ vs. $\tilde m$ plane. 
\begin{figure}[h] 
   \centering
   \includegraphics[width=12cm]{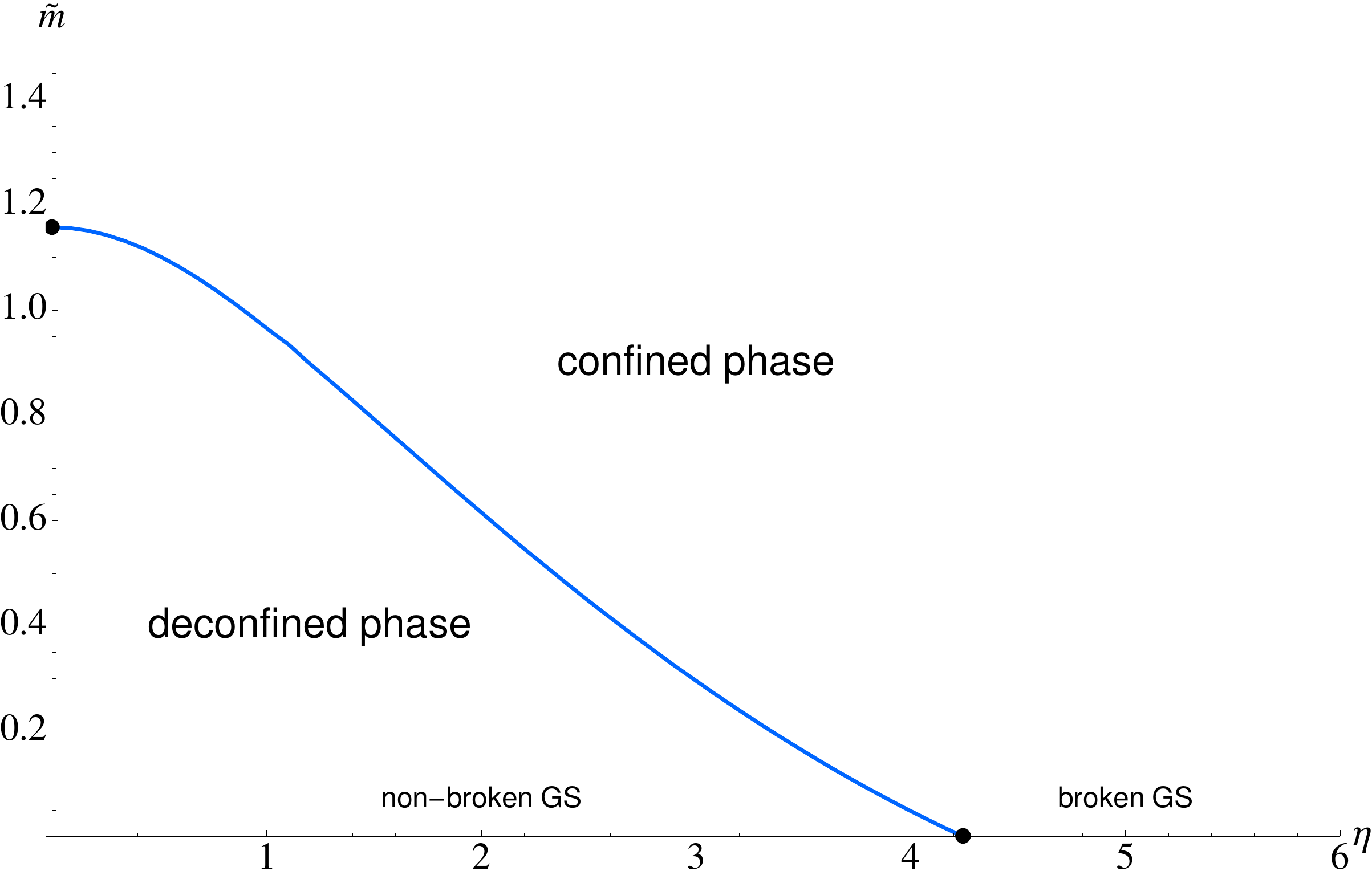}
 
   \caption{\small A phase diagram of the theory. The blue curve separates the confined and deconfined phases of the theory. There is a first order phase transition across the critical curve. For $\eta < \eta_{\rm{cr}}\approx 4.25$ at zero bare mass the theory has a non-broken global SO(3) symmetry, while for $\eta>\eta_{\rm{cr}}$ the global symmetry is spontaneously broken and the theory has a mass-gap.}
   \label{fig:fig3}
\end{figure}
The resulting phase diagram is presented in figure \ref{fig:fig3}. The finite enclosed area corresponds to Black hole embeddings that have a qusi-normal mode excitations and hence correspond to the deconfined phase of the theory. The rest of the phase space correspond to Minkowski embeddings which are characterized by discrete normal excitations that are interpreted as meson-like bound states of the dual gauge theory. This is the confined phase of the theory. Across the critical curve the theory undergoes a first order  confinement/deconfinement phase transition. Note that for $\eta > \eta_{\rm{cr}}\approx 4.25$ the phase transition disappears and the theory is in the confined phase. If we consider the horizontal axes of the diagram ($\tilde m =0$), the confinement/deconfinement phase transition is a ``chiral" phase transition. For $\eta<\eta_{\rm{cr}}$ the theory has an SO(3) global symmetry and no mass-gap, while for $\eta>\eta_{\rm{cr}}$ the theory is in a spontaneously broken phase with finite mass-gap. There are also Goldstone modes associated to the spontaneous symmetry breaking which we shall study in details in Section 4 when we focus on the meson spectrum of the theory. Qualitatively the phase diagram of the theory is the same as the one for the $1+3$ dimensional case studied in \cite{Erdmenger:2007bn,Albash:2007bk}. 

\begin{figure}[h] 
   \centering
   \includegraphics[width=12cm]{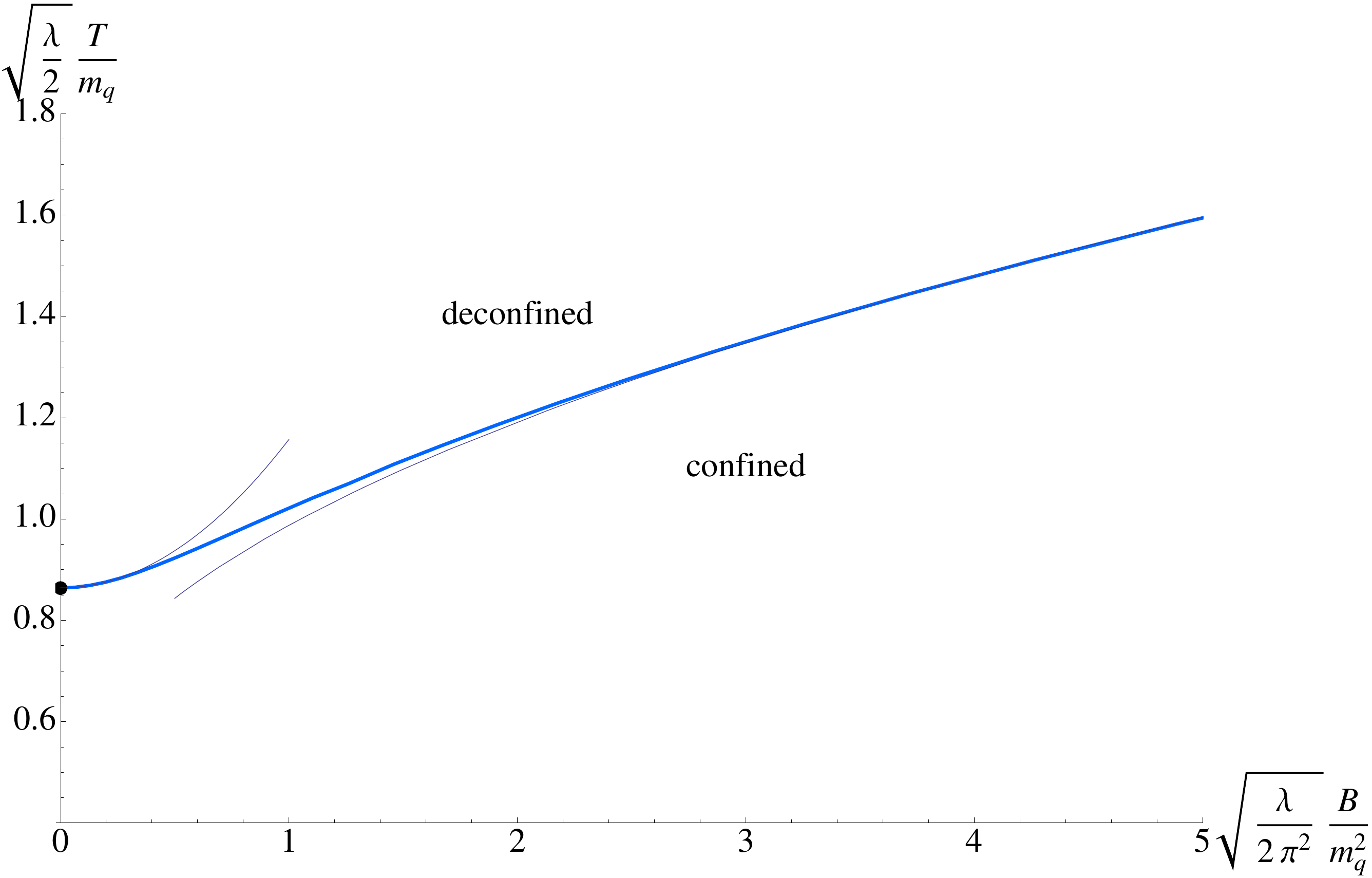}
 
   \caption{\small A phase diagram of the theory in $T$ vs. $B$ coordinates. The fitting curve for small $B$ corresponds to $\sim B^2$ fit and the fitting curve for large magnetic field represents a $\sim\sqrt{B}$ fit.}
   \label{fig:fig4}
\end{figure}

To facilitate comparison to phase diagrams obtained via non-holographic techniques it is useful to represent our phase diagram in field theory units. To this end note that the definitions of our dimensionless quantities $\tilde m,\eta$ imply the following relations:
\begin{equation}
\frac{1}{\tilde m}=\sqrt{\frac{\lambda}{2}}\frac{T}{m_q};~~~\frac{\eta}{\tilde m^2}=\sqrt{\frac{\lambda}{2\pi^2}}\frac{B}{m_q^2}; \ . \label{relTB}
\end{equation}
Where $\lambda$ is the t'Hooft coupling, $T$ is the temperature and $B$ and $m_q$ are the magnetic field and the bare mass of the dual gauge theory. The phase diagram of the theory in these coordinates is presented in figure \ref{fig:fig4}.

The fitting curve for large $B/m_q^2$ corresponds to a $\sim \frac{1}{\sqrt{\eta_{\rm{cr}}}}\sqrt{\sqrt{\frac{\lambda}{2\pi^2}}\frac{B}{m_q^2}}$ behaviour. This reflects the existence of a critical ratio of the magntic field and the temperature at witch the phase transition disappears $\eta_{\rm{cr}}$. Indeed expanding $\tilde m(\eta)$ near $\eta_{\rm{cr}}$ and using that $\tilde m(\eta_{\rm{cr}})\equiv 0$ we obtain:
\begin{equation}
\tilde m(\eta)=(\eta-\eta_{\rm{cr}})\tilde m'(\eta_{\rm{cr}})+O((\eta-\eta_{\rm{cr}})^2) \ .
\end{equation}
Therefore to leading order we have:
\begin{equation}
\frac{1}{\tilde m}=\frac{1}{\sqrt{\eta_{\rm{cr}}}}\sqrt{\frac{\eta}{\tilde m^2 }}-\frac{1}{2\eta_{\rm{cr}}\tilde m'(\eta_{\rm{cr}})}\ ,
\end{equation}
which given the relations from equation (\ref{relTB}) is equivalent to the observed:
\begin{equation}
\sqrt{\frac{\lambda}{2}}\frac{T}{m_q}=\frac{1}{\sqrt{\eta_{\rm{cr}}}}\sqrt{\sqrt{\frac{\lambda}{2\pi^2}}\frac{B}{m_q^2}}+\rm{const};~~~\rm{const}=-\frac{1}{2\eta_{\rm{cr}}\tilde m'(\eta_{\rm{cr}})};
\end{equation}
behaviour for large  $B/m_q^2$. On the other side the fitting curve for small $B/m_q^2$ in figure \ref{fig:fig4} corresponds to a $\sim (\sqrt{\frac{\lambda}{2\pi^2}}\frac{B}{m_q^2})^2$ behaviour. This simply reflects the fact that $\tilde I_{\rm{D5}}$ is a function of even powers of $\eta$ as evident from the definition in equation (\ref{D5dimles}).

It is interesting to compare our phase diagram to similar phase diagrams for $1+2$ dimensional field theories in both external magnetic field and finite temperature. It is somewhat intriguing that the phase diagram of the $1+2$ dimensional Gross-Navie model studied in refs \cite{{Klimenko:1991he},{Klimenko:1992ch}} has the same qualitative structure. It would be interesting to use alternative non-perturbative techniques to study the phase diagram of the defect field theory holographically dual to our set up and compare with the result obtained via the AdS/CFT correspondence. We leave such studies for future investigations.

\subsection{The order parameter}
In this subsection we focus on the zero bare mass case when the confinement/deconfinement phase transition is also a spontaneous symmetry breaking phase transition. The quantity that we are interested in is the fundamental condensate of the theory which serves as an order parameter of the transition. In particular we are interested in the temperature dependence of the condensate at zero bare mass.

Let us consider again the zero temperature case with no external field. As mentioned above at zero bare mass the theory is conformal and has a global $SO(3)_R \times SO(3)$ symmetry. If we introduce external magnetic field \cite{Filev:2009xp} the conformal symmetry is broken and the theory has meson-like excitations. Furthermore the theory develops a negative fundamental condensate that breaks the $SO(3)$ part of the global symmetry down to a $U(1)$ symmetry. There are also zero modes identified with the goldstone bosons of the spontaneously broken global symmetry. On the other side if we turn on a finite temperature the adjoined degrees of freedom are in a deconfined phase. Our holographic set up now describes pure Yang-Mills plasma with some dilute fundamental matter. At low temperatures the fundamental matter is still in the confined phase and is better described as a gas of mesons. However, at sufficiently high temperature the alternative deconfined phase becomes the stable one and the theory undergoes a first order phase transition. The mesons are ``melted" and the dynamically generated mass of the fundamental hypermultiplet fields as well as their condensate vanish. In the holographic set up this is reflected by the fact that the $\tilde l\equiv 0$ embedding of the D5-brane becomes the stable one. This suggests that the global symmetry of the theory is restored in the deconfined phase and we can reffer to the transition as a ``chiral" transition. 

\begin{figure}[h] 
   \centering
   \includegraphics[width=12cm]{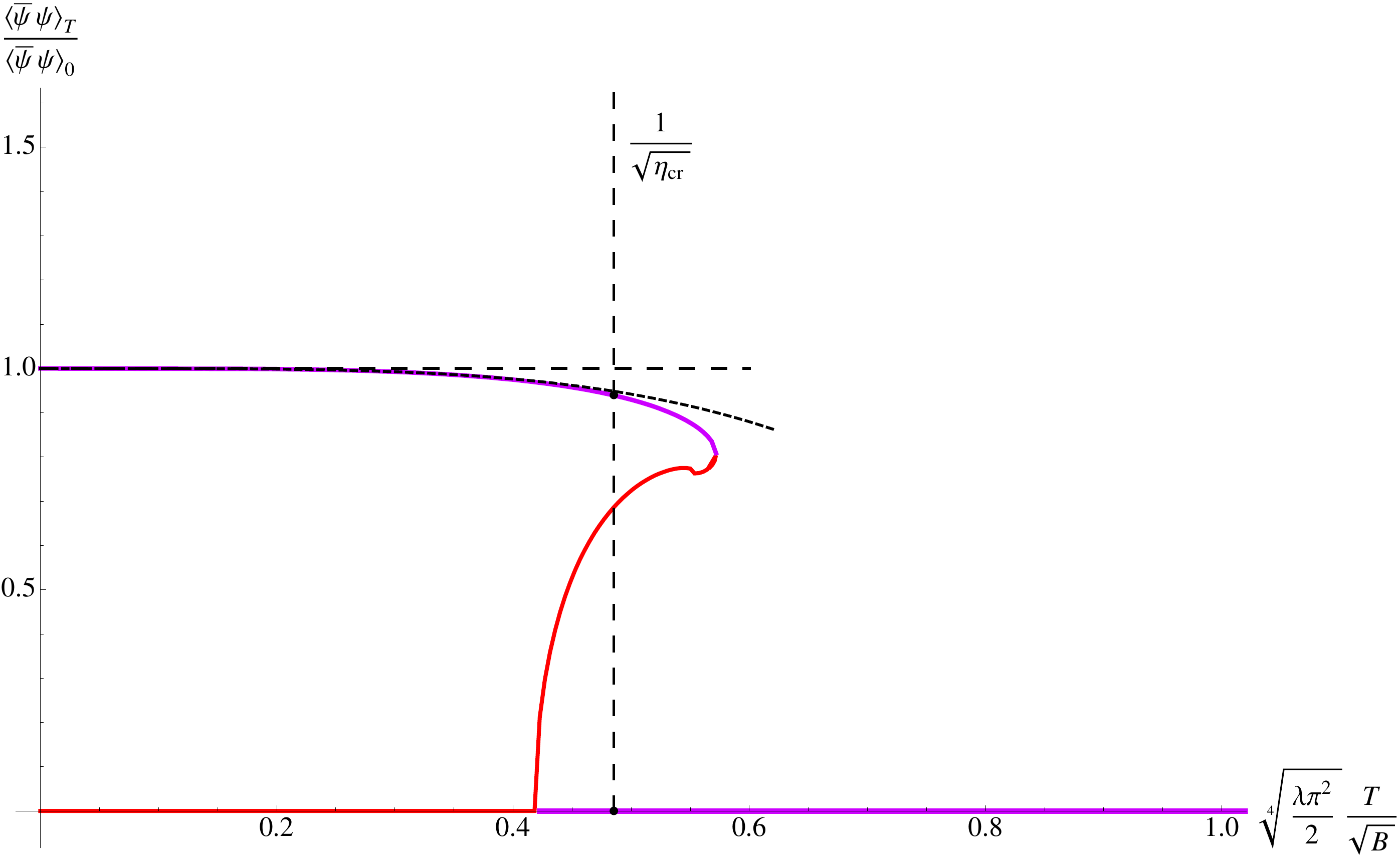}
  \caption{\small A plot of $\langle\bar\psi\psi\rangle_{T}/\langle\bar\psi\psi\rangle_{0}$ versus $T/\sqrt{B}$. The vertical dashe line corresponds to the critical temperature at which the ``chiral"  transition takes place. The purple colour corresponds to either stable or metastable phases while the red colour represents unstable phases with ``tachyonic" excitations. The fitting curve at small temperatures represents  the $1 -{\rm{const}}T^4$ dependence from equation (\ref{depcont}).}
   \label{fig:ordpar}
\end{figure}

The physical picture described above can be visualized if one generates a plot of the temperature dependence of the fundamental condensate (for a fixed magnetic field). More precisely we generate a plot of the ratio of the condensate at finite temperature and the condensate at zero temperature $\langle\bar\psi\psi\rangle_{T}/\langle\bar\psi\psi\rangle_{0}$ versus the ratio of temperature and the square root of the magnetic field $T/\sqrt{B}$. To this end we use that according to equations (\ref{conddef}) and (\ref{bezrazm}) we have the relations:
\begin{equation}
\frac{\langle\bar\psi\psi\rangle_{T}}{B}\propto \frac{1}{\sqrt{\eta}}\tilde c;~~~\frac{1}{\sqrt{\eta}}=\left(\frac{\lambda\pi^2}{2}\right)^{1/4}\frac{T}{\sqrt{B}};\label{nonfant}
\end{equation}
Note also that the ratio ${\langle\bar\psi\psi\rangle_{0}}/{B}$ is obtained in the $\eta\to\infty$ limit in equation (\ref{nonfant}). The resulting plot is presented in figure \ref{fig:ordpar}.

The upper purple curve corresponds to the confined symmetry breaking phase, the colour represents that this phase is ``tachyon" free. (As we shall show in Section 4). The states on the left-hand side of the vertical dashed line are thermodynamically favoured and correspond to a stable phase of the theory. The states on the right-hand side of the critical curve are metastable states. The red curve with negative slope corresponds to unstable phase of the theory which poses a tachyon in the spectrum. It is also unstable from thermodynamic point of view. The horizontal red curve corresponds to a symmetric deconfined phase of the theory which is unstable for temperatures below the critical value (represented by the vertical dashed line). The purple horizontal line corresponds to a symmetric deconfined phase of the theory which is ``tachyon" free. The states on the right-hand side of the vertical dashed line are the thermodynamically favoured ones and correspond to a stable phase of the theory. The states of the horizontal purple line that are on left-hand side of the critical line correspond to a metastable phase of theory and can be reached by supercooling. The fitting curve at low temperatures represents the asymptotic relation:
\begin{equation}
\frac{\langle\bar\psi\psi\rangle_{T}}{\langle\bar\psi\psi\rangle_{0}}=1-{\rm{const}}\left(\frac{\lambda\pi^2}{2}\right)\frac{T^4}{B^2};~~~~~~{\rm{const}}\approx 0.929; \label{depcont}
\end{equation}
This relation can be obtained by expanding the equation of motion of the D5-brane embeddings in powers of $1/\eta$ and solving perturbatively for the leading corrections of the D5-brane embedding. The perturbative solution  is correction near the zero temperature solution obtained in \cite{Filev:2009xp}. It would be nice to derive equation (\ref{depcont}) from field theory side. One can also study this relation in the context of chiral perturbation theory. We leave such studies for future investigations.

\subsection{The entropy denisty}

In order to calculate the entropy density of the theory and verify its positivity we can use equation (\ref{Free-energy}) for the free energy density. The entropy is given given by:
\begin{eqnarray}
S&=&-\left(\frac{\partial F}{\partial T}\right)_H=-\pi R^2\frac{\partial F}{\partial b}=-N_f \frac{\mu_5}{g_s}4\pi^2R^2b^2\left(3\tilde I_{\rm{D5}}+b\frac{\partial\tilde I_{\rm{D5}}}{\partial\tilde m}\frac{\partial\tilde m}{\partial b}+b\frac{\partial\tilde I_{\rm{D5}}}{\partial \eta^2}\frac{\partial\eta^2}{\partial b}\right)\label{entropy}\\
&=&-N_f \frac{\mu_5}{g_s}4\pi^2R^2b^2\left(3\tilde I_{\rm{D5}}+\tilde m\tilde c-4\frac{\partial\tilde I_{\rm{D5}}}{\partial \eta^2}\eta^2\right)=N_f \frac{\mu_5}{g_s}4\pi^2R^2b^2\tilde S(\tilde m,\eta^2);\nonumber\ ,
\end{eqnarray}
where we have used that $\tilde c=-\partial_{\tilde m}\tilde I_{\rm{D5}}$. We have also defined the dimensionless quantity $\tilde S(\tilde m,\eta^2)$ which is convenient to study numerically.

 It is instructive to calculate the entropy of the $\tilde l(\tilde\rho)\equiv 0$ embedding first. That is the entropy at zero bare mass in the deconfined phase with non-broken global symmetry characterized by vanishing condensate. It turns out that in this case the entropy can be obtained in closed form. The result is:
\begin{equation}
\tilde S(0,\eta^2)=\sqrt{1+\eta^2}\label{dimlentm0}\ .
\end{equation}
In field theory units the entropy is:
\begin{equation}
S(m_q=0,T,B)=\sqrt{\lambda/2}N_fN_c\sqrt{T^4+\frac{2B^2}{\lambda\pi^2}}\ .
\end{equation}
Note that at large temperature as $T\to\infty$ (or equivalently weak magnetic field) the entropy grows as $\propto T^2$ as one would expect for a $1+2$ dimensional conformal field theory. However the limit of zero temperature (at fixed magnetic field) doesn't lead to vanishing entropy. This apparently strange behaviour is rectified if one takes into account the phase diagram of the theory. Indeed, at fixed magnetic field the limit of zero temperature suggests that $\eta\to\infty$ which is clearly above the critical value of $\eta_{cr}$ and the stable phase of the theory is the confined phase with spontaneously broken global symmetry. Therefore for very small temperature the entropy at zero bare mass should be evaluated at the state characterized by both vanishing bare mass and (the smallest) negative condensate. The intermediate picture has a first order phase transition at $\eta=\eta_{\rm{cr}}$ with a finite jump of the entropy corresponding to the latent heat of the confinement /deconfinement phase transition. At zero bare mass this transition is also a ``chiral" phase transition. In fact at small (and zero bare mass) the jump of the entropy is enhanced because the dynamical mass generation associated to the ``chiral" phase transition freezes some part of the light degrees of freedom (in the symmetry breaking phase). The latent heat is further increased (relative to the zero magnetic field case) because of the increased diamagnetic response of the theory in the deconfined phase (look at subsection 3.4). 

\begin{figure}[h] 
   \centering
   \includegraphics[width=12cm]{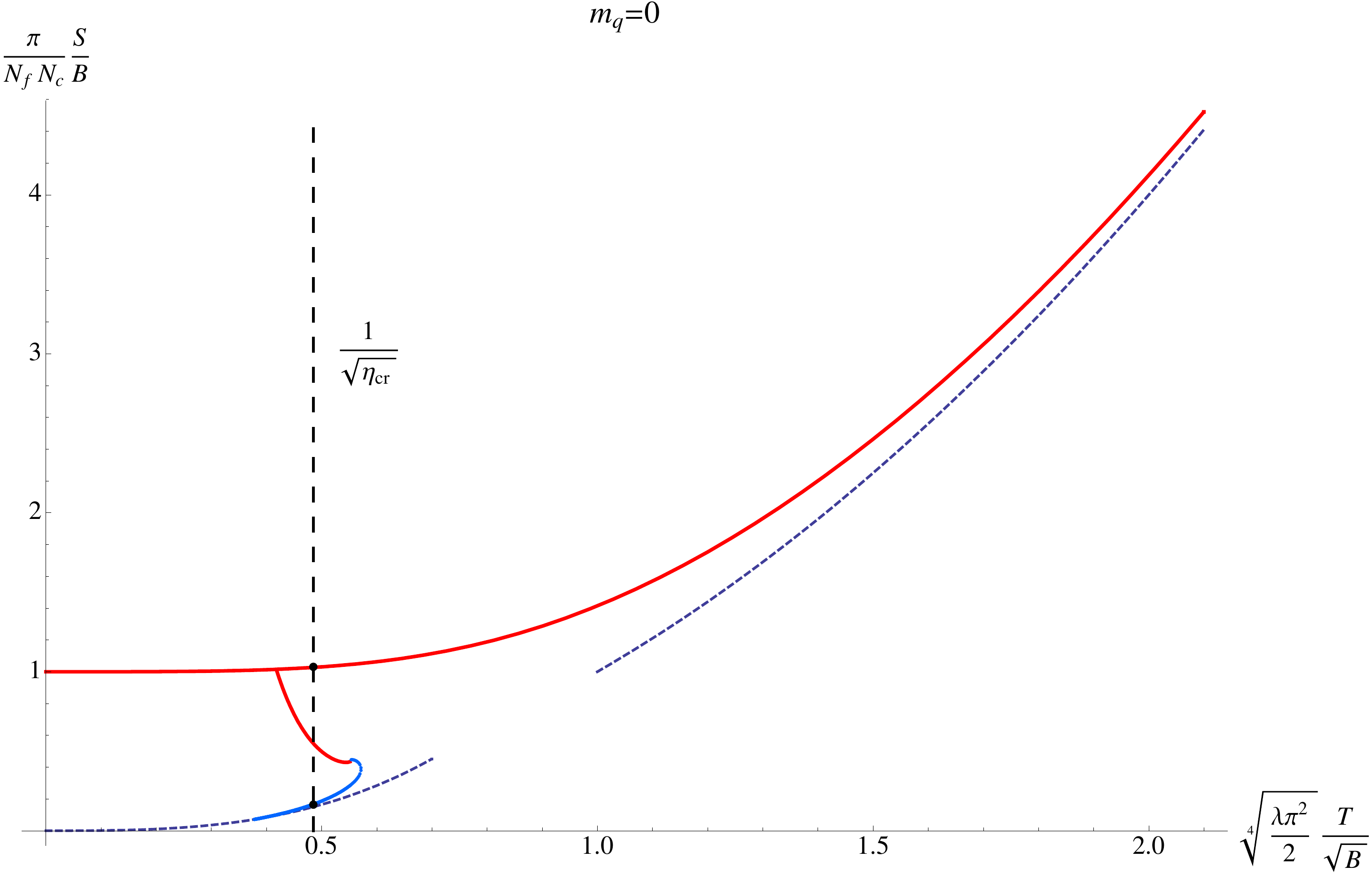}
 
   \caption{\small A plot of the entropy for zero bare mass $m_q=0$. The upper red curve is generated using equation (\ref{dimlentm0}) for $\tilde S(0,\eta)$. The red curve with negative slope is generated numerically and corresponds to an unstable deconfined phase with broken global symmetry. The blue curve corresponds to a confined phase with broken global symmetry. The fitting curve at large temperature corresponds to $~T^2$ dependence as expected for $1+2$ dimensional conformal theory. The fitting curve for small temperature corresponds to $~T^3$ behaviour. There is a first order ``chiral" phase transition at $\eta=\eta_{cr}$ (the vertical dashed line) characterized by a finite positive jump of the entropy denisty. Interestingly the ratio of the entropy immediately before and after the phase transition is equal to $2\pi$. 
}
\label{fig:fig5}
\end{figure}

To generate a plot of the entropy $S$ versus the temperature $T$ at fixed magnetic field $B$ and bare mass $m_q$, we use the following relations:
\begin{equation}
\frac{1}{\sqrt{\eta}}=\left(\frac{\lambda\pi^2}{2}\right)^{1/4}\frac{T}{\sqrt{B}};~~~\frac{\pi}{N_f N_c} \frac{S}{B}=\frac{1}{\eta}\tilde S(\tilde m,\eta);\label{relation}
\end{equation}
Therefore to evaluate the entropy at zero bare mass we need to generate a plot of $\frac{1}{\eta}\tilde S(0,\eta)$ versus $\frac{1}{\sqrt{\eta}}$. The resulting plot is presented in figure \ref{fig:fig5}. 
The red curve in the figure corresponds to the deconfined phase of the theory. As on can see near the phase transition the entropy is a multi-valued function of the temperature. The upper smooth red curve corresponds to a phase with vanishing condensate and non-broken $SO(3)$ symmetry. It is generated using equation (\ref{dimlentm0}) for $\tilde S(0,\eta)$ . The branch with negative slope corresponds to a unstable phase characterized by non-zero condensate and is generated numerically scanning through the corresponding Black hole D5-brane embeddings. The lowest lying blue curve corresponds to a confined phase in which the global symmetry is spontaneously broken. It is generated numerically and corresponds to Minkowski D5-brane embeddings. At $\eta=\eta_{\rm{cr}}$ there is a first order phase transition (represented by the vertical dashed line in the figure). One can see that there is a positive jump of the entropy as going from the confined to the deconfined phase. The fitting curve at high temperature corresponds to $T^2$ behaviour, as expected from a $1+2$ dimensional conformal field theory \cite{Mateos:2007vn} (the effect of the external magnetic field is suppressed in this limit). At low temperature one can show that the entropy vanish as $T^3$. The coefficient of proportionality can be calculated numerically.

\begin{figure}[h] 
   \centering
   \includegraphics[width=8cm]{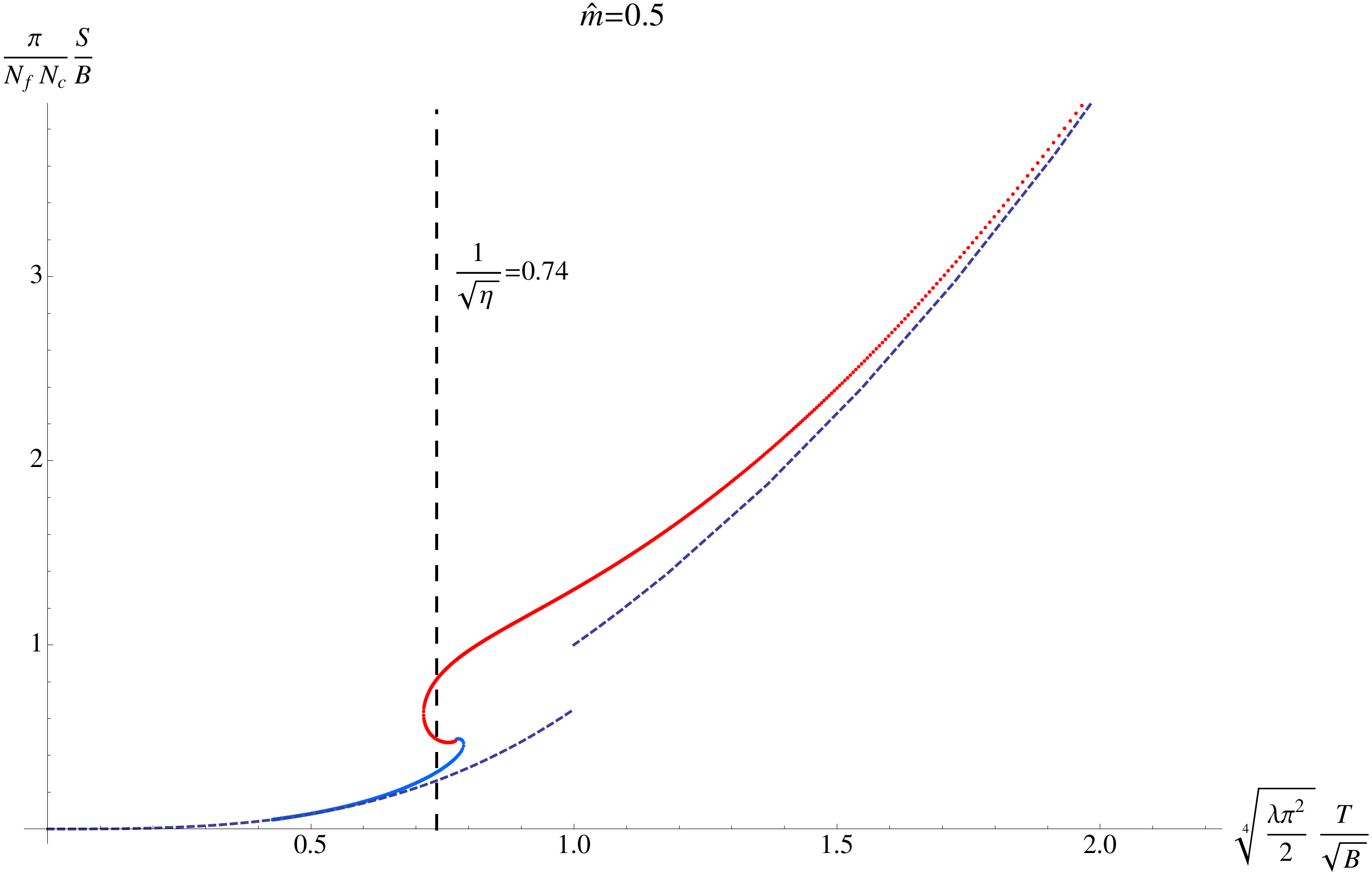}
 \hspace{.5cm}
   \includegraphics[width=8cm]{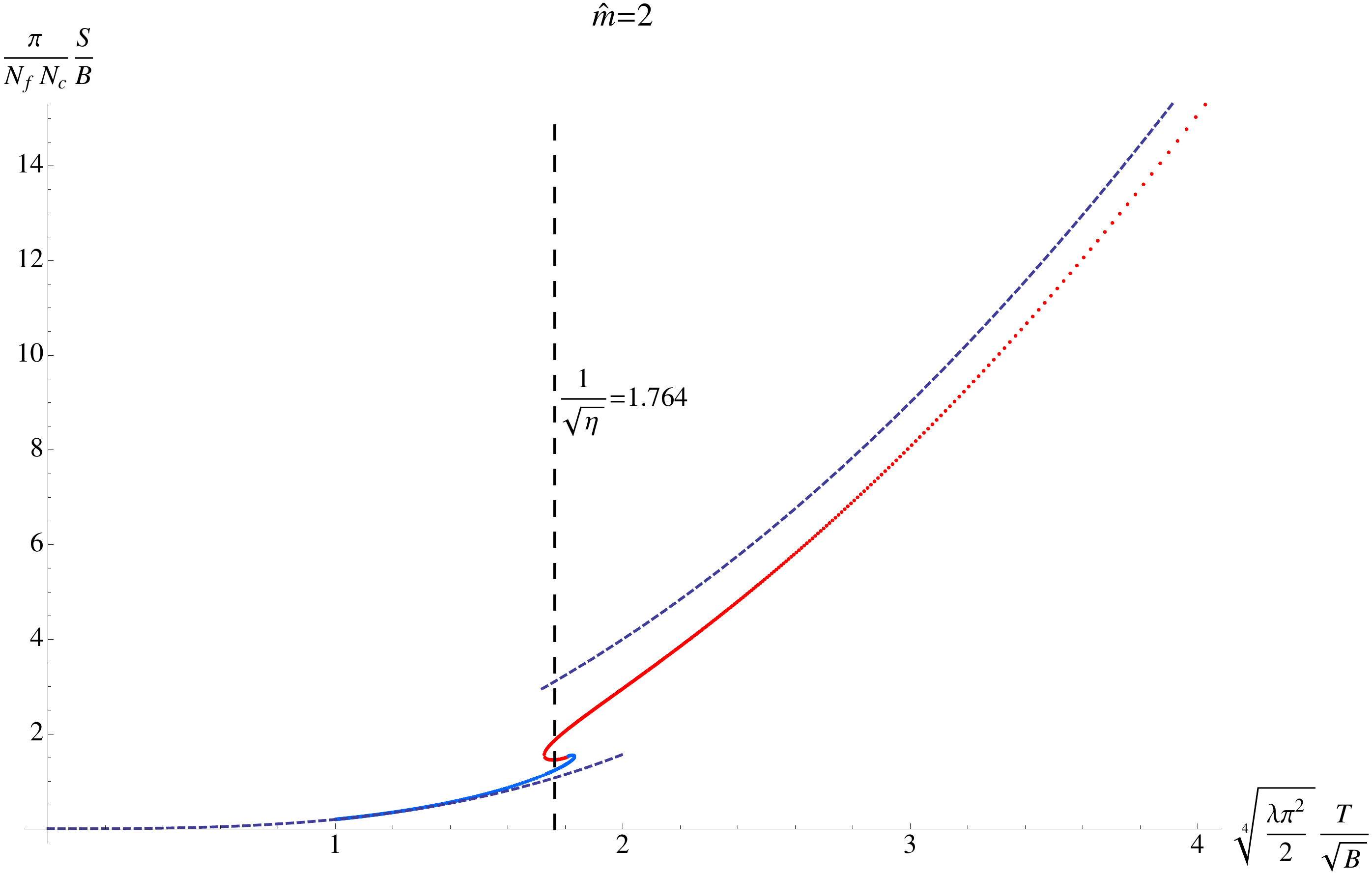} 
   \caption{\small Plots of entropy versus temperature for $\hat m=0.5,2$. The qualitative behaviour is similar to that from figure \ref{fig:fig5}. Major difference is that the deconfined phase with non-broken global symmetry does not exist for temperatures smaller than the temperature at which the negative slope deconfined phase appears.}
   \label{fig:fig6}
\end{figure}

For generic bare masses the entropy can be studied in two different regimes: the regime of fixed ratio of magnetic field and temperature (fixed $\eta$) and the regime of fixed ratio of bare mass and external magnetic field (fixed $m_q/\sqrt{B}$). The former regime is technically more convenient to study (the quantity $\tilde I_{\rm{D5}}$ depends explicitly on $\eta$). This is also the regime studied in refs. \cite{Albash:2007bk,Erdmenger:2007bn}. However in this regime the slope of the entropy $S$ versus temperature $T$ plot is not proportional to the specific heat at fixed magnetic field $c_{B}$, because the magnetic field varies with temperature (in order to keep $\eta\propto B/T^2$ fixed). This is why we study the entropy at fixed ratio of the bare mass and the magnetic field. This is also a natural generalization of the study of the zero bare mass case and as we are going to see the qualitative behaviour remains unchanged. To perform the study we use the dimensionless variables defined in equation $(\ref{relation})$. The parameter that we keep fixed is:
\begin{equation}
\hat m=\frac{\tilde m}{\sqrt{\eta}}=m/R\sqrt{H}=\left(\frac{2\pi^2}{\lambda}\right)^{1/4}\frac{m_q}{\sqrt{B}}\ .\label{hatm}
\end{equation}
The resulting plots for $\hat m=0.5,2$ are presented in figure \ref{fig:fig6}. 
Similarly to the zero bare mass case at high temperatures the entropy grows as $T^2$ which is the dependence expected for $1+2$ dimensional conformal field theory at finite temperature. For intermediate temperatures there is a finite jump of the entropy associated to the confinement/deconfinement phase trasition of the fundamental matter. At low temperatures the entropy goes to zero as $T^3$. The coefficient of proportionality depends on the ratio $m_q/\sqrt{B}$ and can be calculated numerically. Note that near the phase transition the entropy is a multi-valued function of the temperature. The multiple phases regime of the theory exists for a finite interval of temperatures and at the boundaries of this interval the specific heat $c_{\rm{B}}$ diverges. A more detailed study of the entropy function in the unstable phase would reveal that the negative slope curve has more complicated structure. Namely, there is a double logarithmic structure and the red and blue curves join in the centre of that spiral \cite{Mateos:2006nu}. We shall not study further this property of the theory. 

To determine the critical value of $1/\sqrt{\eta}$ we have used the phase diagram from figure \ref{fig:fig4}. Alternatively we could have used the {\it equal areas law} \cite{Albash:2007bk}, because the area below the curves in figure \ref{fig:fig6} is proportional to the free energy of the system.

\subsection{The magnetization}

In the statistical ensemble that we consider, that is fixed temperature $T$ and fixed magnetic field $B$, there is another natural quantity of interest. This is the magnetization of the theory given by:
\begin{eqnarray}
&&M=-\left({\frac{\partial F}{\partial B}}\right)_{T}=-8\pi^2\alpha' N_f\frac{\mu_5}{g_s}b^3\left(\frac{\partial\tilde I_{\rm{D5}}}{\partial\eta}\right)_{\tilde m}\frac{R^2}{b^2}=\label{magnetization}\\
&&=-8\pi^2\alpha' N_f\frac{\mu_5}{g_s}R^2b\int\limits_{\tilde\rho_{\rm{min}}}^{\infty}d\tilde\rho\frac {2\eta \tilde\rho^2\left (-1 + 4 \tilde r^4 \right)\sqrt {1 +\tilde l'^2}} {
 \tilde r^ 2\sqrt {4\tilde r^4 + 1}\sqrt {\left (1 + 4\tilde r^4 \right)^2 +16\tilde r^4\eta^2}}\equiv8\pi^2\alpha' N_f\frac{\mu_5}{g_s}R^2b\tilde M\ .\nonumber
\end{eqnarray}
Here we have defined the dimensionless quantity $\tilde M$. In filed theory units the last expression in equation (\ref {magnetization}) can be rewritten as:
\begin{equation} 
M=\frac{N_fN_c}{\pi}T\tilde M\ .
\end{equation}
Thus the quantity $\tilde M$ is convenient for studies at fixed non-zero temperature. In order to incorporate the zero temperature case it is convenient to define:
\begin{equation}
M=\frac{2^{1/4}N_fN_c}{\lambda^{1/4}\pi^{3/2}}\sqrt{B}\frac{1}{\sqrt{\eta}}\tilde M\equiv \frac{2^{1/4}N_fN_c}{\lambda^{1/4}\pi^{3/2}}\sqrt{B}\hat M\ ,\label{dimlesmag}
\end{equation}
where we have used the first equation from (\ref{relation}). The dimensionless quantity $\hat M$ is given by:
\begin{equation}
\hat M=-\int\limits_{\hat\rho_{\rm{min}}}^{\infty}d\hat\rho\frac {2\hat\rho^2\left (-1/\eta^2 + 4 \hat r^4 \right)\sqrt {1 +\hat l'^2}} {
 \hat r^ 2\sqrt {4\hat r^4 + 1/\eta^2}\sqrt {\left (1/\eta^2 + 4\hat r^4 \right)^2 +16\hat r^4}}\ .\label{intexpMhat}
\end{equation}
Note that we have introduced a new set of dimensionless variables:
\begin{equation}
\hat\rho=\rho/R\sqrt{H};~~~\hat l=l/R\sqrt{H};\label{dimlesvarmag}
\end{equation}
convenient to study the theory at a fixed magnetic field. 
It is instructive to study first the magnetization at zero temperature.

\subsubsection{Magnetization at zero temperature}
\begin{figure}[h] 
   \centering
   \includegraphics[width=12cm]{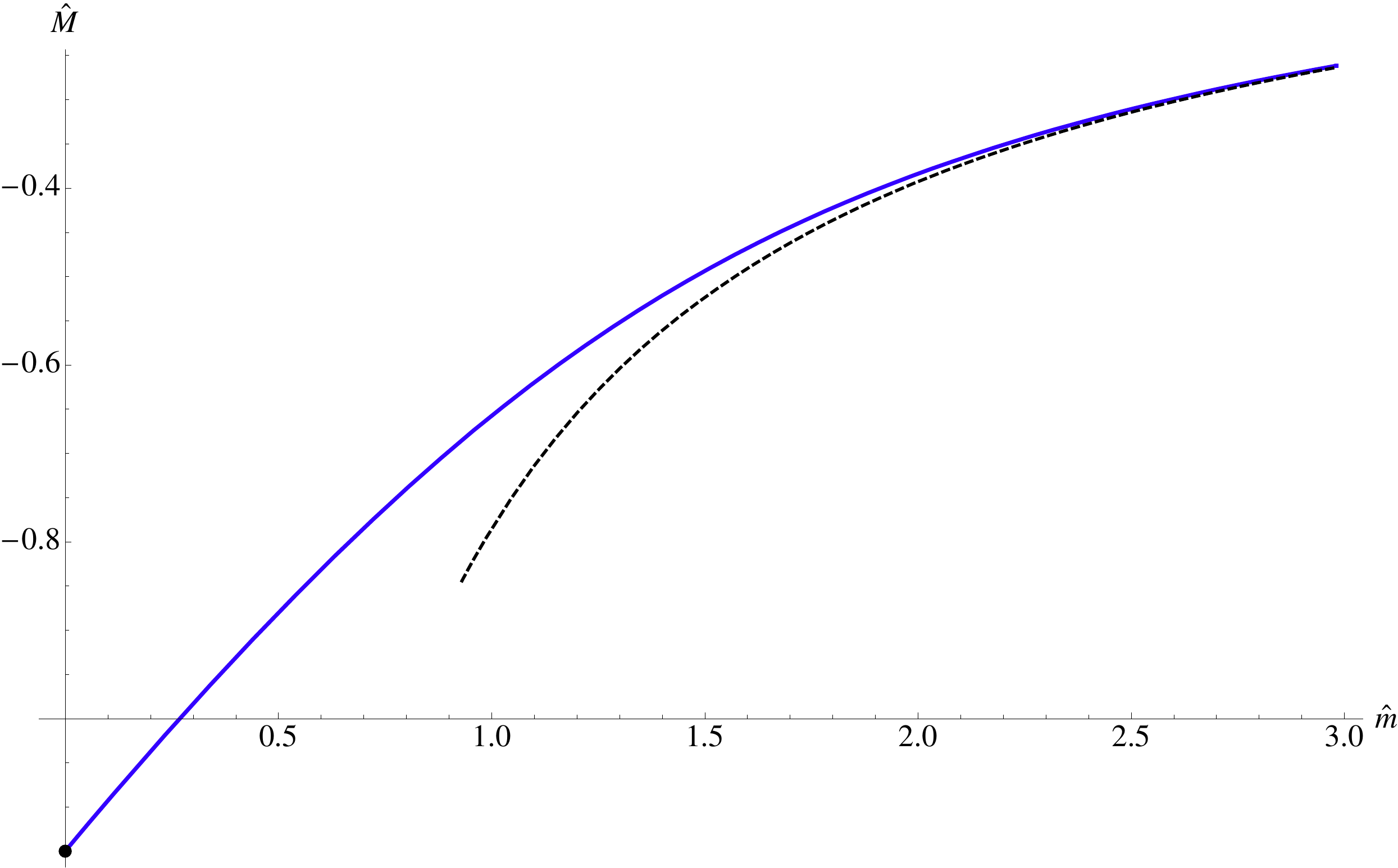}
 
   \caption{\small A plot of $\hat M_0\propto M_0/\sqrt{B}$ versus $\hat m\propto m_q/\sqrt{B}$. The dashed curve corresponds to equation (\ref{larghatm}) one can see the good agreement for large bare masses. At zero bare mass the magnetization has a finite negative value. The finite positive slope at small $\hat m$ suggests finite negative susceptibility.}
\label{fig:fig7}
\end{figure}

The zero temperature case corresponds to the limit $\eta\to\infty$ in equation (\ref{intexpMhat}). The expression for the magnetization $\hat M$ is given by:
\begin{equation}
\hat M_0=-\int\limits_{\hat\rho_{\rm{min}}}^{\infty}d\hat\rho\frac {\hat\rho^2\sqrt {1 +\hat l'^2}} {
 \hat r^ 2\sqrt {1 +\hat r^4}}\ .\label{intexpMhat0}
\end{equation}
Here $\hat l(\hat\rho)$ is a solution to the equation of motion for the D5 brane probing a pure $AdS_5\times S^5$ space. We shall be interested only on the stable branch of the theory. For large bare masses $\hat m\gg1$ one can use the results from ref. \cite{Filev:2009xp} for the asymptotic behaviour of the condensate:
\begin{equation} 
\hat c=-\frac{\pi}{8\hat m^2}+\dots
\end{equation}
to show that the asymptotic behaviour of the magnetization should be:
\begin{equation}
\hat M_0=-\frac{\pi}{4\hat m}\ .\label{larghatm}
\end{equation}
after using equations (\ref{hatm}) and (\ref{dimlesmag}) we obtain:
\begin{equation}
M_0(B)=-\frac{N_fN_c}{4\pi}\frac{B}{m_q}+\dots
\end{equation}
Therefore at weak magnetic field (large bare mass) the meson gas has diamagnetic properties with negative susceptibility:
\begin{equation}
 \chi_0=\frac{\partial M_0}{\partial B}=-\frac{N_fN_c}{4\pi}\frac{1}{m_q}+\dots\ .
\end{equation} 
For large magnetic field the magnetization can be obtained numerically. The resulting plot is presented in figure \ref{fig:fig7}.
As one can see equation (\ref{larghatm}) is a good approximation for large bare mass (weak magnetic field) while at zero bare mass the magnetization has a non-vanishing negative value. Furthermore the finite positive slope at small $\hat m$ suggests finite negative susceptibility. This confirms that at zero temperature the meson gas is in a diamagnetic phase. One may expect that at finite temperature the diamagnetic properties of the meson gas would not depend strongly on the temperature. However at high enough temperature a confinement/deconfinement phase transition takes place and the theory is in a conductive phase. It would be interesting to study the effect that this transition has on the magnetic properties of the matter.
\subsubsection{Magnetization at finite temperature}
It is instructive to study first the magnetization at zero bare mass $m_q=0$. At zero bare mass the theory can be in either confined or deconfined phase depending on the ratio of the magnetic field and the temperature squared. This is controlled by the parameter $\eta$. For small values of $\eta$ the theory is in the deconfined phase and has a non-broken SU(3) global symmetry. The corresponding D5 brane embedding is the $l\equiv0$ one. For large enough temperatures (weak magnetic field) $\tilde M$ behaves as:
\begin{equation}
\tilde M=-\eta+O(\eta^2)\ .\label{highTM} 
\end{equation}
This suggests the following expression for the magnetization $M$ in field theory units:
\begin{equation}
M=-\frac{2^{1/2}N_fN_c}{\lambda^{1/2}\pi^2}\frac{B}{T}\left(1+O\left(\frac{B}{T^2}\right)\right)\ .\label{largeTM}
\end{equation}
Clearly the magnetization is negative and the leading contribution to the magnetic susceptibility is:
\begin{equation}
\chi=-\frac{2^{1/2}N_fN_c}{\lambda^{1/2}\pi^2}\frac{1}{T}\ .\label{susc}
\end{equation}
Therefore the deconfined phase of the theory is a diamagnetic phase. Furthermore the diamagnetic response depends strongly on the temperature and goes to zero as the temperature approaches infinity. Curiously equation (\ref{susc}) is similar to the Curie's law $\chi\propto 1/T$ for a paramagnetic but with the wrong sign. This is also the most simple expression for the magnetic susceptibility in $1+2$ dimensions based on dimensional analysis only. Such a behaviour is to be expected since at high temperatures conformality is restored.

\begin{figure}[h] 
   \centering
   \includegraphics[width=12cm]{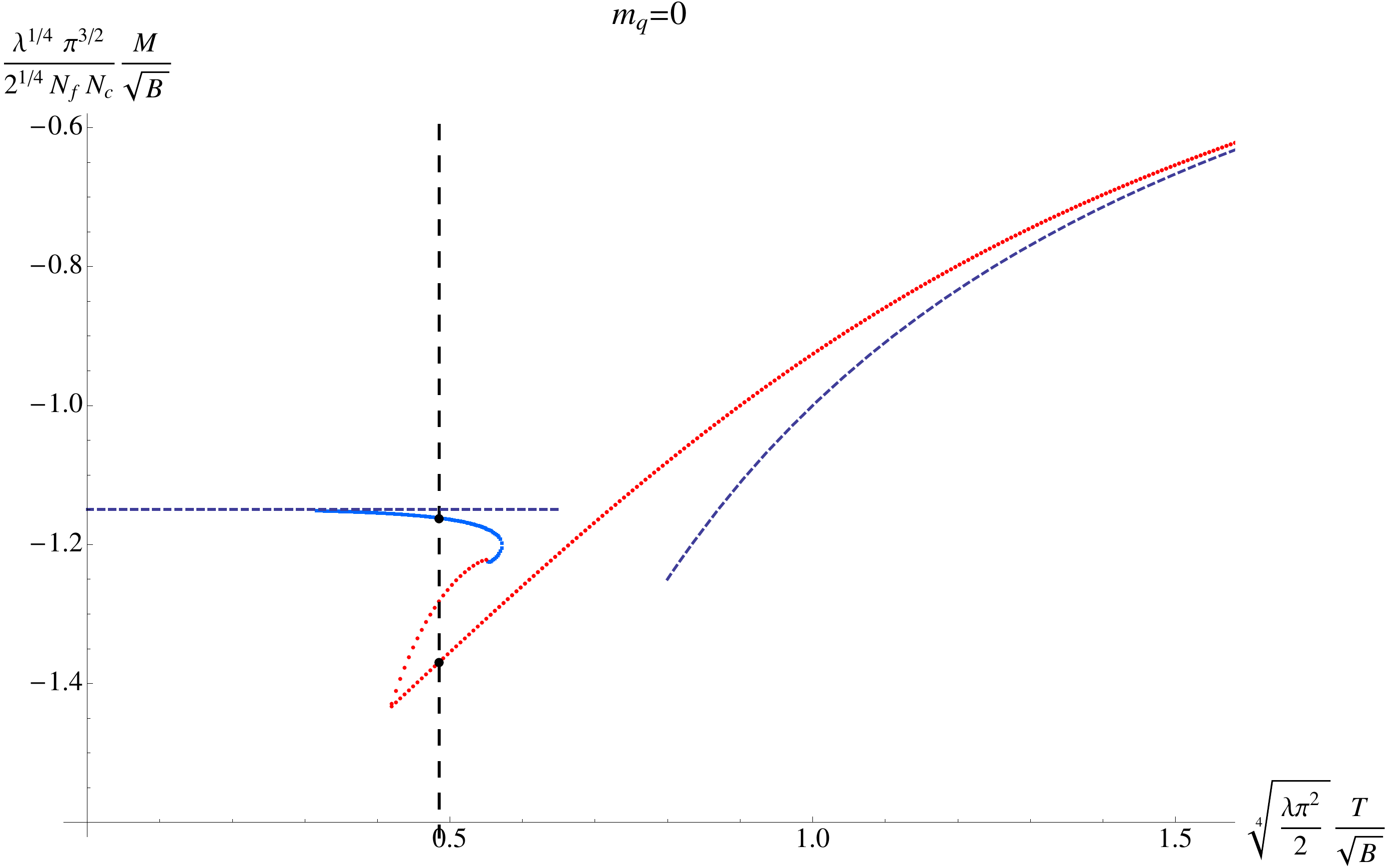}
 
   \caption{\small A plot of $\hat M$ versus $1/\sqrt{\eta}$. The axes are labeled in field theory units. The blue curve corresponds to the confined global symmetry breaking phase of the theory. The dashed horizontal line corresponds to $\hat M_0|_{\hat m=0}$. The dashed vertical line denotes the first order confinement/deconfinement phase transition. The fitting curve at high $T$ corresponds to equation (\ref{largeTM}).}
\label{fig:fig8}
\end{figure}

For strong magnetic field the magnetization can be obtained numerically. It is convenient to use the parameter $\hat M\propto M/\sqrt{B}$ defined in equation(\ref{dimlesmag}). A plot of $\hat M$ versus $1/\sqrt{\eta}$ is presented in figure \ref{fig:fig8}. Note that the axes are labeled in field theory units.
The blue curve for low temperature corresponds to the confined global symmetry breaking phase of the theory. The dashed horizontal line corresponds to $\hat M_0|_{\hat m=0}$. As one can see the magnetization is negative and the diamagnetic response of the theory varies slightly with the temperature. At sufficiently high temperature (the vertical line in the figure) the theory undergoes a first order confinement/deconfinement phase transition, which is also a ``chiral" phase transition. As one can see there is a significant jump of the magnetization suggesting much stronger diamagnetic response of the theory in the deconfined phase. This is expected because the confined phase is also a conductive phase. Upon further increase of the temperature the magnetization approaches the behaviour described by equation (\ref{largeTM}) (the fitting curve for high $T$). One can see that in the deconfined phase the magnetization depends strongly on the temperature. One explanation is that the conductivity drops as the temperature is increased. 

Another interesting feature is that for the minimal value of $1/\sqrt{\eta}$ for which the deconfined exists there is a kink and slope has a finite jump. This is in contrast to the point corresponding to the maximum value of $1/\sqrt{\eta}$ for which the confined phase exists. The slope diverges at this point and the blue curve bends smoothly. It seems that this is specific for the $m_q=0$ case.

\begin{figure}[h] 
\centering
\includegraphics[width=8cm]{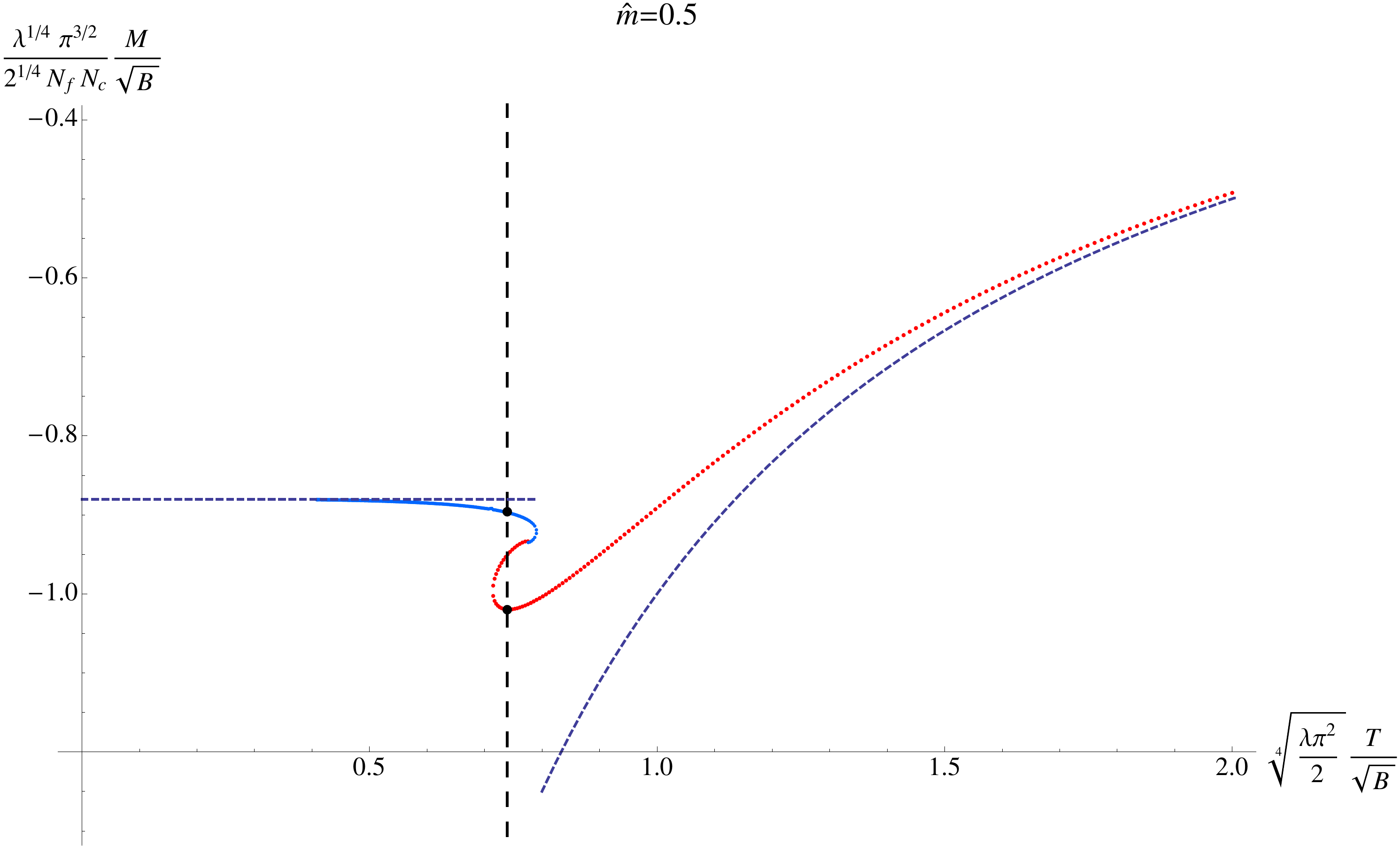}
\hspace{.5cm}
\includegraphics[width=8cm]{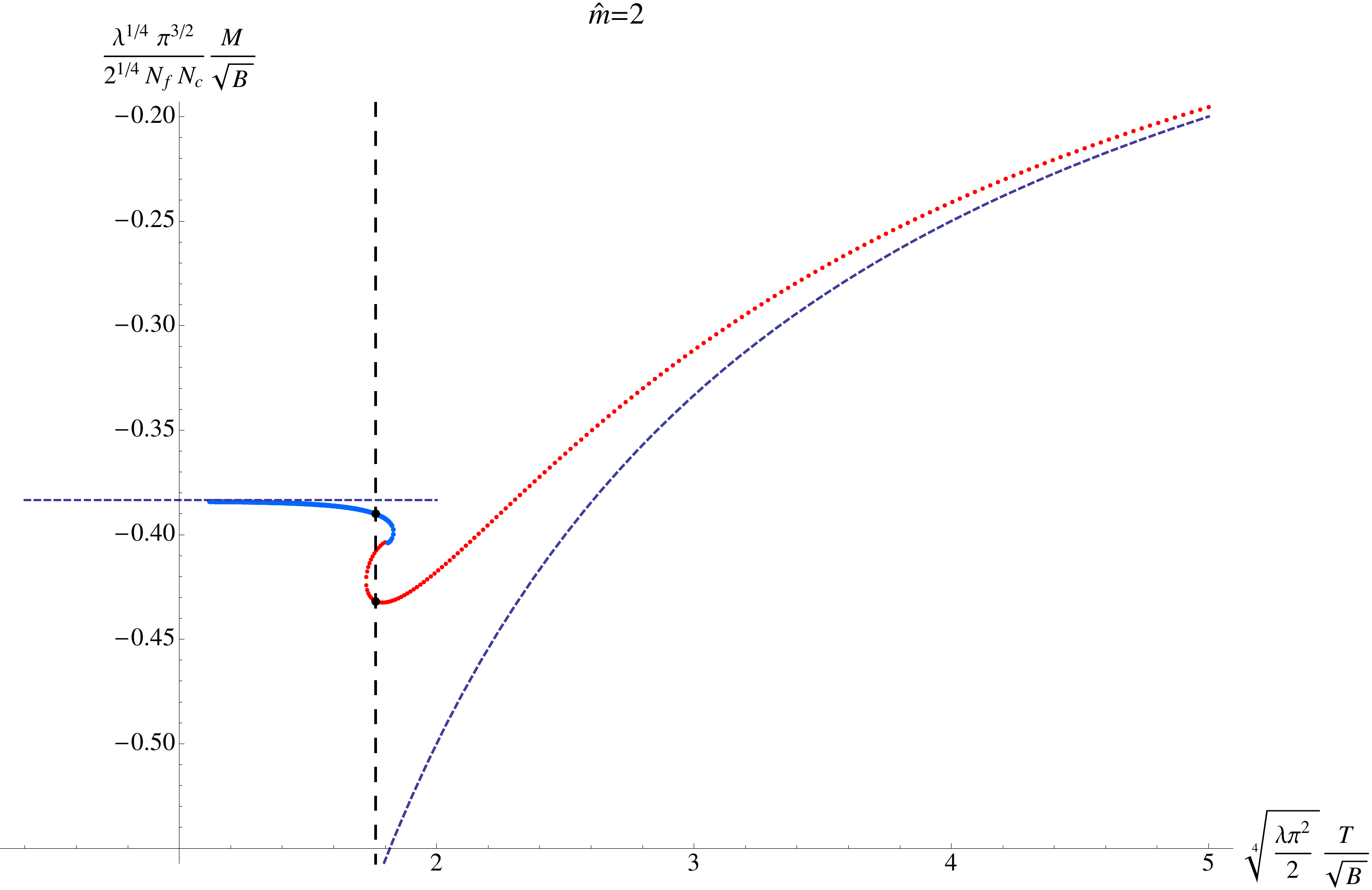}
\caption{\small Plots of $\hat M$ versus $1/\sqrt{\eta}$ for $\hat m=0.5, 2$. The blue curves correspond to the confined phase and the horizontal dashed lines correspond to $\hat M_0|_{\hat m=0.5}$ and $\hat M_0|_{\hat m=2}$ correspondingly. The vertical lines correspond to $\eta_{\rm{cr}}$ for which the confinemet/deconfinement phase transition takes place. The red curves describe the magnetization in the deconfined phase of the theory and the fitting curves correspond to equation (\ref{largeTM}). }
\label{fig:fig9}
\end{figure}

The case of generic values of $\hat m$ can also be studied numerically. Plots of $\hat M$ versus $1/\sqrt{\eta}$ for $\hat m=0.5, 2$ are presented in figure \ref{fig:fig9}. 
As one may expect the magnetization is negative and for temperatures below the critical one the theory is in a confined phase (the blue curves in the figure). One can see that the magnetization varies slowly with temperature and approaches very fast the values for vanishing temperature ($\hat M_0|_{\hat m=0.5}$ and $\hat M_0|_{\hat m=2}$) represented by the horizontal dashed lines. At the confinement/deconfinement transition there is a jump of the magnetization. It is interesting that the relative jump of the magnetization seems to decrease as $\hat m$ increase ($\hat m=0,0.5,2$). For high temperature the theory is in a deconfined phase and the magnetization approaches the behviour described by equation (\ref{largeTM}). Unlike the $\hat m=0$ case there is no kink at the point where the deconfined phase cease existence. 

\section{Meson Spectrum}

In this section we study the light meson spectrum of the theory. The holographic prescription for obtaining the spectrum of meson-like excitations is to study semi-classical fluctuations of the probe flavour branes. Let us consider a small excitation of the probe brane propagating along the radial direction of the geometry. Since the geometry is asymptotically AdS space-time the excitations would bounce at infinity without dissipation. On the other side the geometry has a horizon in the bulk and there are two possibilities. If the D5-brane has a shrinking $S^2$ and close above the horizon (Minkowski type of embeddings) the excitations would bounce and as a result the D5-brane would have normal modes corresponding to standing waves. Another possibility is that the D5-brane reaches the horizon before the $S^2$ shrinks (Black hole embeddings) in this case the excitation fall into the horizon and dissipate. The corresponding semiclassical excitations are quasi-normal modes satisfying ingoing boundary conditions at the horizon. 

The interpretation on gauge theory side of the correspondence is that the normal modes of Minkowski embeddings correspond to the meson spectrum of the confined phase, while Black hole embeddings correspond to the deconfined phase of theory and their quasi-normal modes describe mesons melting in the Yang-Mills plasma \cite{Hoyos:2006gb}. The real components of the quasi-normal modes correspond to the energy of the melting meson and the imaginary component correspond to the relaxation time (if it is negative).

A study of the semiclassical excitations of the theory is of a particular interest when we have multiple phases because it verifies the thermodynamic analysis of the stability of different phases (existence or non-existence of tachyonic modes). It is also important when we have a spontaneous breaking of a continuous symmetry because of the existence of Goldstone modes and related phenomena. 

The meson spectrum of the holographic gauge theory corresponding to the D3/D5 set up has been extensively studied in the literature. In refs.~\cite{Arean:2006pk,Myers:2006qr} the spectrum of the supersymmetric zero temperature case has been investigated. In ref.~\cite{Filev:2009xp} an external magnetic field has been introduced leading to a non-supersymmetric field theory exhibiting spontaneous symmetry breaking. The corresponding meson spectrum has been investigated with a focus on the pseudo-Golstone modes. A non-relativistic dispersion relation of the goldstone modes has been revealed and various phenomenological relations verified. However analysis of the stability based on the massive meson modes has not been performed. In this paper we shall perform such studies in the more general case when a finite temperature is turned on. 

In order to study the light meson spectrum of the theory we look for
the quadratic fluctuations of the D5--brane embedding along the
transverse directions parametrized by $l,\psi,\phi$. To this end we
expand:
\begin{equation}
l=\bar l+2\pi\alpha'\delta l;~~~\psi=2\pi\alpha'\delta\psi;~~~\phi=2\pi\alpha'\delta\phi\, ,
\end{equation}
in the action (\ref{DBI2}) and leave only terms of order
$(2\pi\alpha')^2$. Note that fluctuations of the $U(1)$ gauge field
$F_{\alpha\beta}$ of the D5--brane will also contribute to the
expansion. There is also an additional contribution from the
Wess-Zumino term of the D5--brane's action:
\begin{equation}
S_{\rm{WZ}}=N_f\mu_5\int\limits_{{\cal M}_6}\sum_p[C_{p}\wedge e^{\cal F}];~~~{\cal F}=B+2\pi\alpha'F\, .
\end{equation}

For the anzatz that we are considering, the relevant term is:
\begin{equation}
S_{\rm{WZ}}=N_f\mu_5\int\limits_{{\cal M}_6}B\wedge P[\tilde C_{4}]\ ,
\end{equation}
where $P[\tilde C_4]$ is the pull-back of the magnetic dual, $\tilde
C_{4}$, to the background $C_{4}$ R-R form. For the particular
parameterization of $S_5$ considered here, it is given by:
\begin{equation}
\tilde C_4=\frac{1}{g_s}\frac{4\rho^2l^2}{(\rho^2+l^2)^3}R^4\sin\psi(l d\rho- \rho dl)\wedge d\Omega_2\wedge d\phi\ .
\end{equation}
After some long but straightforward calculations we get the following
action for the quadratic fluctuations along $l$:
\begin{eqnarray}
  &&{\cal L}_{ll}^{(2)}\propto\frac{1}{2}\sqrt{-E}G_{ll}\frac{S^{\alpha\beta}}{1+l'^2}\partial_{\alpha}\delta l\partial_{\beta}\delta l+\frac{1}{2}\left[\partial_{l}^2\sqrt{-E}-\frac{d}{d\rho}\left(\frac{l'}{1+l'^2}\partial_l\sqrt{-E}\right)\right]\delta l^2\ ,\\ 
  &&{\cal L}_{lF}^{(2)}\propto\frac{\sqrt{-E}}{1+l'^2}(\partial_l J^{12}-\partial_{r}J^{12}l')F_{21}\delta l\nonumber\ ,\\
  &&{\cal L}_{FF}^{(2)}\propto\frac{1}{4}\sqrt{-E}S^{\alpha\beta}S^{\gamma\lambda}F_{\beta\gamma}F_{\alpha\lambda}\nonumber\ ,
\end{eqnarray}
and along $\phi$ and $\psi$:
\begin{eqnarray}
&&{\cal L}_{\psi\psi,\phi\phi}^{(2)}\propto\frac{1}{2}\sqrt{-E}S^{\alpha\beta}(G_{\psi\psi}\partial_{\alpha}\delta\psi\partial_{\beta}\delta\psi+G_{\phi\phi}\partial_{\alpha}\delta\phi\partial_{\beta}\delta\phi)\ \label{qvfl},\\
&&{\cal L}_{\psi\phi}^{(2)}\propto (\cos\alpha) PH\delta\psi\partial_0\delta\phi\nonumber \ .
\end{eqnarray}
Here $E_{\alpha\beta}$ is the pull-back of the generalized metric on
the classical D5--brane embedding:
\begin{equation}
E_{\alpha\beta}=\partial_{\alpha}\bar X^{\mu}\partial_{\beta}\bar X^{\nu}(G_{\mu\nu}+B_{\mu\nu})\, ,
\end{equation}
and we have defined $S^{\alpha\beta}$ and $J^{\alpha\beta}$ as the
symmetric and anti-symmetric elements of the inverse generalized
metric $E^{\alpha\beta}$:
\begin{equation}
E^{\alpha\beta}=S^{\alpha\beta}+J^{\alpha\beta}\ .
\end{equation}
The functions $P$ and $g(\rho)\equiv\sqrt{-E}/\cos\alpha$ are given by:
\begin{eqnarray}
&&g(\rho)=\rho^2\left(1-\frac{b^4}{4r^4}\right)\left(1+\frac{b^4}{4r^4}\right)^{1/2}\left(1+\frac{16R^4H^2r^4}{(4r^4+b^4)^2}\right)^{1/2}\sqrt{1+l'^2}\ ,\\
&&P=\frac{4R^4\rho^2l^2}{(\rho^2+l^2)^3}(\rho l'-l)\ .
\end{eqnarray}

\subsection{Fluctuations along l.}
 We first focus our attention to the fluctuations along $l$. Note that there is no symmetry associated to  translations along $l$ and we do not expect the appearance of Goldstone bosons. However the absence or presence of tachyons in the spectrum of $\delta l$ can give is information about the stability of the phase that we are studying. 
 
 The equation of motion derived from the quadratic action
(\ref{qvfl}) is  the following:
\begin{eqnarray}
\partial_\rho\left(\frac{g(\rho)\delta l'}{(1+l'^2)^2}\right)+\frac{4g(\rho)R^4}{(1+l'^2)(4r^4+b^4)} \Box\delta l+\frac{g(\rho)}{1+l'^2}\frac{\Delta_{(2)}}{\rho^2}\delta l-\left[\partial_{l}^2g(\rho)-\frac{d}{d\rho}\left(\frac{l'\partial_l g(\rho)}{1+l'^2}\right)\right]\delta l=0\label{fluctL0}
\end{eqnarray} 
 where
 \begin{equation}
 \Box=\frac{-\partial_0^2}{1-\frac{16b^4r^4}{(4r^4+b^4)^2}}+\frac{\partial_1^2+\partial_2^2}{1+\frac{16R^4H^2r^4}{(4r^4+b^4)^2}};\label{Box}
 \end{equation}
 and $\Delta_{(2)}$ is the spherical laplacian. Note that the modified operator of d'Alembert reflects the breaking of the $SO(1,2)$ Lorentz symmetry down to a $SO(2)$ symmetry. 
 
Let us focus on the meson spectrum of the various phases at zero bare mass. In particular let us verify that the symmetry breaking phase corresponding to a confined phase of the theory (the upper purple curve in figure \ref{fig:ordpar} ) is stable or metastable and has no tachyon modes. To this end we consider a plane-wave ansatz:
\begin{equation}
\delta\tilde l=\zeta(\rho)e^{-i\omega t+i\vec k.\vec x}\ .
 \end{equation}
Since we are interested only on the energy of the excitations we will further restrict our anzats to $\vec k=0$. the equation of motion can be written as:
\begin{equation}
\partial_{\tilde\rho}\left(\frac{\tilde g(\tilde \rho)\zeta'}{(1+\tilde l'^2)^2}\right)+\frac{4\tilde g(\tilde \rho)(4\tilde r^4+1)}{(1+\tilde l'^2)(4\tilde r^4-1)^2}\tilde\omega^2\zeta-\left[\partial_{\tilde l}^2\tilde g(\tilde\rho)-\frac{d}{d\tilde\rho}\left(\frac{\tilde l'\partial_{\tilde l} \tilde g(\tilde\rho)}{1+\tilde l'^2}\right)\right]\zeta=0\label{fluctL}\ .
\end{equation}
Where we have rewritten the equation in dimensionless variables and defined $\tilde\omega=R^2\omega/b$.
\subsubsection{Normal modes.}
Next we consider the Minkowski embeddings that asymptote to zero separation $\tilde m=0$ at infinity and solve numerically the equation of motion (\ref{fluctL}) imposing the following boundary conditions at $\tilde\rho=\epsilon=10^{-6}$:
\begin{equation}
\zeta(\epsilon)=1;~~~\zeta'(\epsilon)=0;
\end{equation}
the spectrum is obtained by requiring regularity of the solution at infinity. 

\begin{figure}[h] 
\centering
\includegraphics[width=12cm]{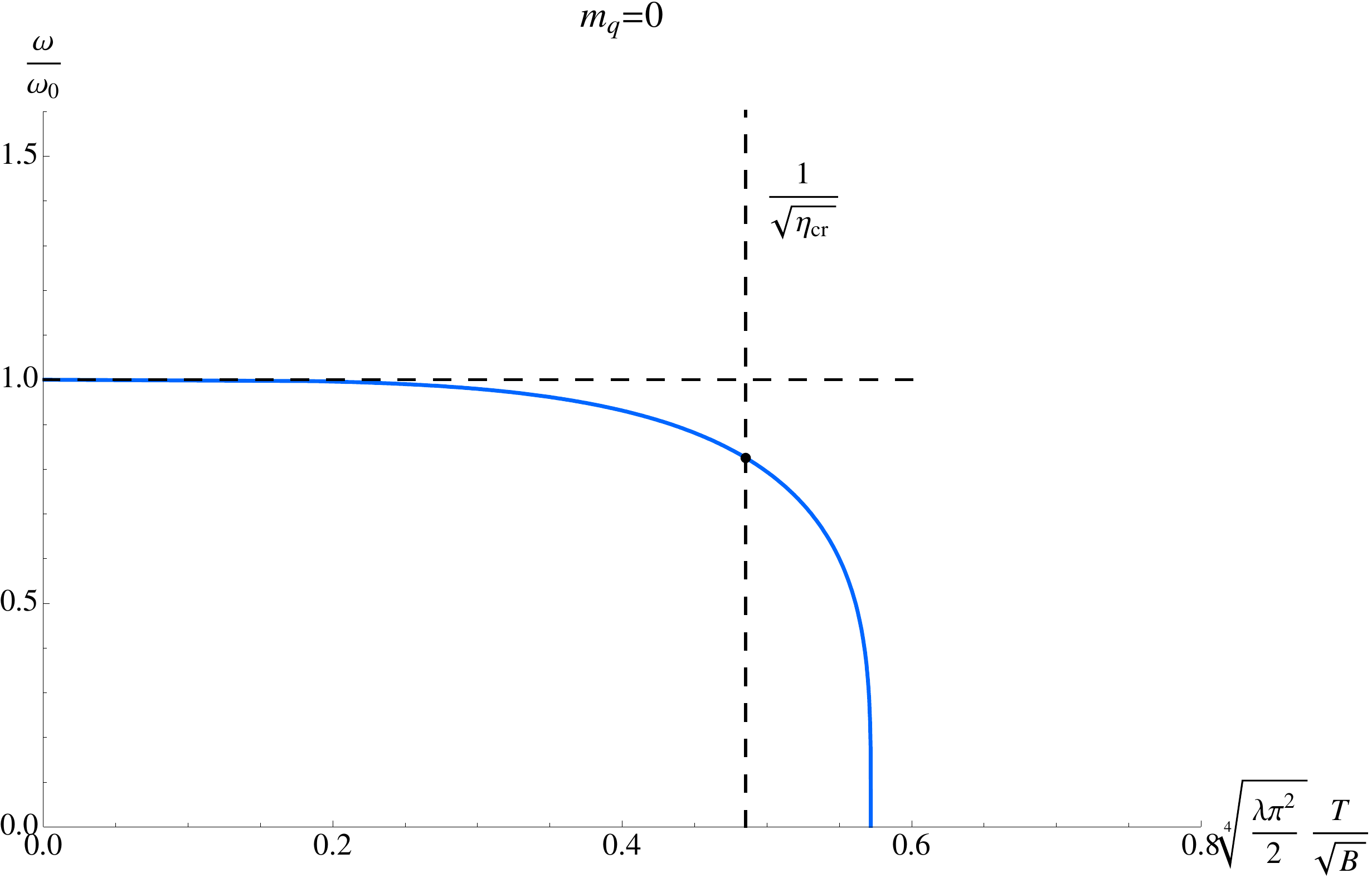}
\caption{\small A plot of the spectrum of the ground state of the symmetry breaking phase. The spectrum remains real even for temperature above the critical temperature, represented by the vertical dashed line. The spectrum drops to zero at the state corresponding to divergent heat capacity.}

\label{fig:fig10}
\end{figure}

To study the ground state of the meson spectrum in the symmetry broken phase and its temperature dependence we generate a plot of $\omega/\omega_0$ versus $T/{\sqrt{B}}$ using that:
\begin{equation}
\frac{\omega}{\sqrt{B}}\propto \frac{1}{\sqrt{\eta}}\tilde\omega
\end{equation}
and that the zero temperature value of the frequency $\omega_0$ corresponds to the $\eta\to\infty$ limit. A plot of of $\omega/\omega_0$ versus $1/\sqrt{\eta}$ is presented in figure \ref{fig:fig10}. 

As one can see for low temperature the ground state is tachyon free and the confined symmetry breaking phase is the stable one. Furthermore even for temperatures above the critical one (the vertical dashed line in the figure) the symmetry breaking phase remains metastable. If we further increase the temperature the energy of the ground state decreases sharply. Remarkably at the maximal temperature for witch confined phase exists the energy of the ground state drops to zero. Note that at this point the heat capacity also diverges as the plot of the entropy from figure \ref{fig:fig5} infers. Note that near the maximal temperature there is an alternative confined phase represented by a red curve in figure \ref{fig:ordpar} and having negative slope in figure \ref{fig:fig5}. A study of the meson spectrum shows that this phase has tachyonic ground state and is unstable. The spectrum is represented by the blue curve in figure \ref{fig:fig11}. 

\subsubsection{Quasi-normal modes}

To study the deconfined phase we need to study the qusi-normal excitations of the probe D5-branes. This suggests imposing ingoing boundary conditions at the horizon of the geometry this is why it is more convenient to use spherical coordinates in the transverse $\IR^6$ subspace. The equations are further simplified if we introduce a new radial coordinate:
\begin{equation}
u^2=\frac{4r^4+b^4}{4r^2}; 
\end{equation}
In the new coordinates the metric of the $AdS_5\times S^5$ black hole is given by:
\begin{equation}
ds^2/\alpha'=-\frac{u^4-b^4}{R^2u^2}dt^2+\frac{u^2}{R^2}d\vec x^2+\frac{R^2u^2}{u^4-b^4}du^2+R^2(d\theta^2+\cos\theta^2d\Omega_2^2+\sin\theta^2\tilde\Omega_2^2)\ .
\end{equation}
note that in these coordinates fluctuations along $l$ corresponds to fluctuations along $\theta$. It is also a straightforward exercise to rewrite the equation of motion for $\delta l$ (\ref{fluctL0}) in terms of $\theta(u)$ and $\delta\theta(u)$. It is convenient to introduce dimensionless coordinates $\tilde u=u/b$. The resulting equation of motion is given by:
\begin{eqnarray}
&&\partial_{\tilde u}\left(\frac{g({\tilde u})({\tilde u}^4-1)\delta\theta'}{(1+\frac{{\tilde u}^4-1}{{\tilde u}^2}\theta'^2)^2\tilde u^2}\right)+\frac{g(\tilde u)}{(1+\frac{\tilde u^4-1}{\tilde u^2}\theta'^2)\tilde u^2}\tilde{\Box} \delta\theta+\frac{g(\tilde u)}{1+\frac{\tilde u^4-1}{\tilde u^2}\theta'^2}\frac{\Delta_{(2)}}{\cos\theta^2}\delta\theta\\
&&-\left[\partial_{\theta}^2g(\tilde u)-\partial_{\tilde u}\left(\frac{\frac{\tilde u^4-1}{\tilde u^2}\theta'}{1+\frac{\tilde u^4-1}{\tilde u^2}\theta'^2}\partial_{\theta}g(\tilde u)\right)\right]\delta\theta=0\ ,\nonumber\label{fluctthet}
\end{eqnarray}
where
\begin{eqnarray}
&&g(\tilde u)=\tilde u^2\cos\theta^2\left(1+\frac{\eta^2}{\tilde u^4}\right)^{1/2}\left(1+\frac{\tilde u^4-1}{\tilde u^2}\theta'^2\right)^{1/2};\\
&&\tilde\Box=\frac{R^4}{b^2}\left[\frac{-\partial_0^2}{1-\frac{1}{\tilde u^4}}+\frac{\partial_1^2+\partial_2^2}{1+\frac{\eta^2}{\tilde u^4}}\right];\nonumber
\end{eqnarray} 

To study the spectrum of the fluctuations we consider an anzats:
\begin{equation}
\delta\theta=e^{-i\omega t}\zeta(\tilde u)\label{anzquaz}
\end{equation}
Note that in general $\omega$ is a complex number. The real part is naturally interpreted as the energy of the excitation, while the imaginary part of $\omega$ is proportional to the relaxation time of the excitation, provided that it is negative ($\rm{Im}(\omega)<0$). Clearly a positive imaginary part would lead to exponentially growing (with time) excitations. The existence of such modes signals instability of the phase under consideration. 

\begin{figure}[h] 
\centering
\includegraphics[width=12cm]{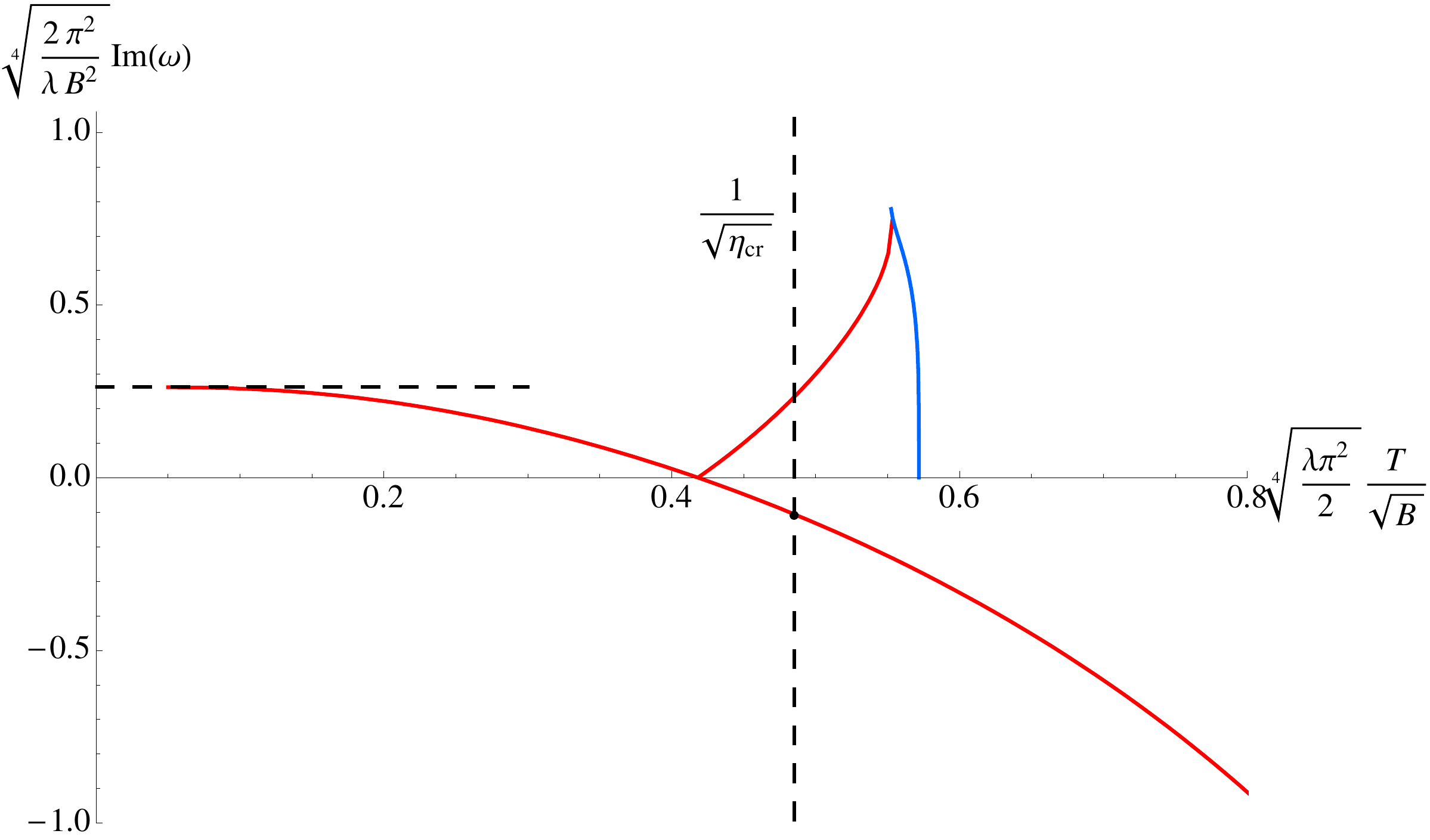}
\caption{\small A plot of the imaginary part of the of the quasi-normal modes (the red curves) and of the tachyonic sector of the normal modes (the blue curve). The vertical dashed line corresponds to the critical temperature at which the confinement/deconfinement phase transition takes place.The horizontal dashed line corresponds to the highest quasi-normal mode at zero temperature.}
\label{fig:fig11}
\end{figure}

To quantize the spectrum we need to impose ingoing boundary conditions at the horizon. To this end we focus on the equation of motion for $\zeta(\tilde u)$ near 
$\tilde u=1$. To leading order we have:
\begin{equation}
\zeta''(\tilde u)+\frac{1}{\tilde u-1}\zeta'(\tilde u)+\frac{\tilde\omega^2}{16(\tilde u-1)^2}\zeta(\tilde u)=0\ ,\label{quasi1}
\end{equation}
where we have defined $\tilde\omega=\frac{R^2}{b}\omega$. The most general solution of equation (\ref{quasi1}) is a linear combination of $(\tilde u-1)^{\pm i\tilde \omega/4}$. However ingoing boundary conditions correspond to the negative sign solution $(\tilde u-1)^{- i\tilde \omega/4}$ (for the choice of signs considered in $(\ref{anzquaz})$). To impose the condition we define:
\begin{equation}
\zeta(\tilde u)=(\tilde u -1)^{-i\tilde\omega/4}S(\tilde u);
\end{equation}
and solve numerically the resulting equation of motion for $S(\tilde u)$ imposing Dirichlet boundary condition at the horizon ($S(1)=1;$). The quasi-normal modes are obtained by selecting regular solutions at infinity. More precisely we require that $|(\tilde u -1)^{-i\tilde\omega/4}S(\tilde u)|\sim 1/\tilde u^2$ as we send $\tilde u\to\infty$.
Next we consider the classical black hole embeddings which have a zero separation at infinity $\tilde m=0$. There are two classes of solutions. Solutions which develop separation in the bulk of the geometry and hence describe deconfined symmetry breaking phase and the $\theta\equiv 0$ solution which has restored symmetry. It turns out that that the spectrum of quasi-normla modes for both type of solutions contain a mode which is purely imaginary and thus particularly convenient to study numerically. A plot of the spectrum is presented in figure \ref{fig:fig11}.
The smooth red curve corresponds to the $\theta\equiv 0$ solution and thus describe a deconfined phase with restored global symmetry. As one can see for large temperatures the imaginary part of the quasi-normal mode is negative and thus the phase is stable. As we lower the temperature we reach the critical temperature represented by the vertical dashed line. At this point a first order phase transition takes place and the thermodynamically favoured phase is the confined symmetry breaking phase. However as one can see from figure \ref{fig:fig11} the symmetric phase remains metastable even for temperatures below the critical one and thus describe states that can be reached by supercooling. If we continue to lower the temperature the imaginary part of the quasi-normal mode becomes positive and the deconfined phase is unstable. Remarkably this happens at the point where the symmetry breaking black hole solutions and the $\theta\equiv0$ solutions meet. In the limit $T\to0$ the spectrum approaches the spectrum for pure $AdS_5\times S^5$ space represented by the horizontal dashed line.

The symmetry breaking deconfined phase is represented by the short red curve with positive slope in figure \ref{fig:fig11} (and negative slope in figure \ref{fig:fig5}). As one can see it has quasi-normal excitations with positive imaginary components and is thus unstable. This fits nicely with the fact that it has a negative heat capacity (negative slope in figure \ref{fig:fig5}). 

Finally the blue curve in figure \ref{fig:fig11} corresponds to Minkowski solutions which have tachyonic meson spectrum and is obtained by solving equation (\ref{fluctL}). This phase is represented by a blue curve in the entropy versus temperature plot (figure \ref{fig:fig5}) and has a negative slope. Clearly it is unstable as both the presence of tachyons and negative heat capacity suggest. Note that the kink in figure \ref{fig:fig11}, where the red and blue curves join, corresponds to embeddings near the critical embedding which separates black hole embeddings and Minkowski embeddings. A more detailed analysis of the theory in this regime would reveal a discrete self-similar behaviour \cite{Mateos:2007vn}. We shell not elaborate further on this property of the set up.

\subsubsection{The zero magnetic field case}

For completeness and to verify the validity of our numerical analysis it is instructive to study in details the spectrum of quasi-normal modes at finite temperature and zero magnetic field. At zero bare mass we can employ the technique used in ref. \cite{Hoyos:2006gb} and solve the corresponding Heun equation. 

To begin with let us write down the equation of motion (\ref{fluctthet}) for the fluctuations of the $\theta\equiv 0$ embedding:
\begin{equation}
\delta\theta''(\tilde u)+\left(\frac{4\tilde u^3}{\tilde u^4-1}-\frac{2\eta^2}{\tilde u(\tilde u^4+\eta^2)}\right)\delta\theta'(\tilde u)+\left(\frac{\tilde\omega^2\tilde u^4}{(\tilde u^4-1)^2}+\frac{2\tilde u^2}{\tilde u^4-1}\right)\delta\theta(\tilde u)=0\label{thet0}
\end{equation}
Next we focus on the $\eta=0$ case and consider the substitution $x=1-1/\tilde u^2$. The equation of motion is written as:
\begin{equation}
\delta\ddot\theta+\frac{1+3(1-x)^2}{2x(1-x)(2-x)}\delta\dot\theta+\left(\frac{\tilde\omega^2}{4x^2(1-x)(2-x)^2}+\frac{2}{4x(1-x)^2(2-x)}\right)\delta\theta=0\label{eqnfluctx}
\end{equation}
Equation (\ref{eqnfluctx}) has four regular singularities $x=0,1,2,\infty$ with exponents $\{-i{\tilde w}/{4},+i{\tilde w}/{4}\}$, $\{1/2,1\}$, $\{-\tilde\omega/4,+\tilde\omega/4\}$, $\{0,0\}$. Upon the change of variables:
\begin{equation} 
\delta\theta(t,x)=e^{-i\omega t}x^{-i\frac{\tilde\omega}{4}}(1-x)(x-2)^{-\frac{\omega}{4}}y(x)
\end{equation}
equation (\ref{eqnfluctx}) takes the standard form of a Heun equation
\begin{equation}
\ddot y(x)+\left(\frac{\gamma}{x}+\frac{\delta}{x-1}+\frac{\epsilon}{x-2}\right)\dot y(x)+\frac{\alpha\beta-Q}{x(x-1)(x-2)}y(x)=0\ ,\label{Heun}
\end{equation}
with parameters:
\begin{eqnarray}
&&\gamma=1-i{\tilde w}/{2};~~\delta=3/2;~~\epsilon=1-\tilde\omega/2;~~Q=3/2-(1/4+i)\tilde\omega-(1/4-i/8)\tilde\omega^2;\\
&&\alpha=3/2-(1/4+i/4)\tilde\omega;~~\beta=1-(1/4+i/4)\tilde\omega;~~\epsilon = \alpha + \beta-\gamma-\delta +1;\nonumber \ .
\end{eqnarray}
Next we look for solutions of equation (\ref{Heun}) satisfying $y(0)=y(1)=1$. This selects solutions of equation (\ref{thet0}) obeying ingoing boundary condition at the horizon ($x=0$) and an appropriate behaviour at infinity ($x=1$). The method that we use is the one employed in ref. \cite{Hoyos:2006gb} for the D3/D7 set up. We consider a Frobenius series near $x=0$ satisfying the recursion relation:
\begin{equation}
a_{n+2}+A_n(\tilde\omega)a_{n+1}+B(\tilde\omega)a_n=0\ ,
\end{equation}
where $A_n(\tilde\omega)$ and $B_n(\tilde\omega)$ are given by:
\begin{eqnarray}
&&A_{n}(\tilde\omega)=-\frac{(n+1)(2\delta+\kappa+3(n+\gamma))+Q}{ 2 (n + 2) (n + 1 + \gamma)}\\
&&B_n(\tilde\omega)=\frac{(n+\alpha)(n+\beta)}{2(n+2)(n+1+\gamma)}
\end{eqnarray}
and $a_0=1$, $a_1=Q/2\gamma$. Next we define a continued fraction \cite{Starinets:2002br}, \cite{Hoyos:2006gb}:
\begin{equation}
r_n=\frac{a_{n+1}}{a_n}=-\frac{B_n(\tilde\omega)}{A_n(\tilde\omega)+r_{n+1}}\ ,
\end{equation}
the convergency condition is given by:
\begin{equation}
r_0=Q/2\gamma\ .
\end{equation}
To obtain approximate expression for $r_0$ we cut the recursive relation at some sufficiently large $n$ ($n=150$ in our case) and use the asymptotic expression for $r_n$:
\begin{equation}
r_n=\frac{1}{2}-\frac{2+\tilde\omega}{4n}+\dots
\end{equation}
Next we solve the resulting algebraic equation for $\tilde\omega$. The spectrum for the first 11 lowest quasi-normal modes is given in table \ref{table1}.
\begin{table}[!htbp]
\hspace*{4ex}
\begin{tabular}{cc}
\parbox[c]{.35\textwidth}{%
\vspace*{-16em}
\begin{tabular}{|c|c|c|} \hline
k&Re$(\tilde\omega)$&Im$(\tilde\omega)$\\ \hline\hline
0&$\pm$1.6906&-1.3268\\\hline
1&$\pm$3.6805&-3.3269\\\hline
2&$\pm$5.6780&-5.3261\\\hline
3&$\pm$7.6769&-7.3258\\\hline
4&$\pm$9.6764&-9.3256\\\hline
5&$\pm$11.676&-11.326\\\hline
6&$\pm$13.676-&13.325\\\hline
7&$\pm$15.676&-15.325\\\hline
8&$\pm$17.676&-17.325\\\hline
9&$\pm$19.675&-19.325\\\hline
10&$\pm$21.675&-21.325\\\hline
\end{tabular}} &
\parbox[t]{.6\textwidth}{\includegraphics[scale=0.75]{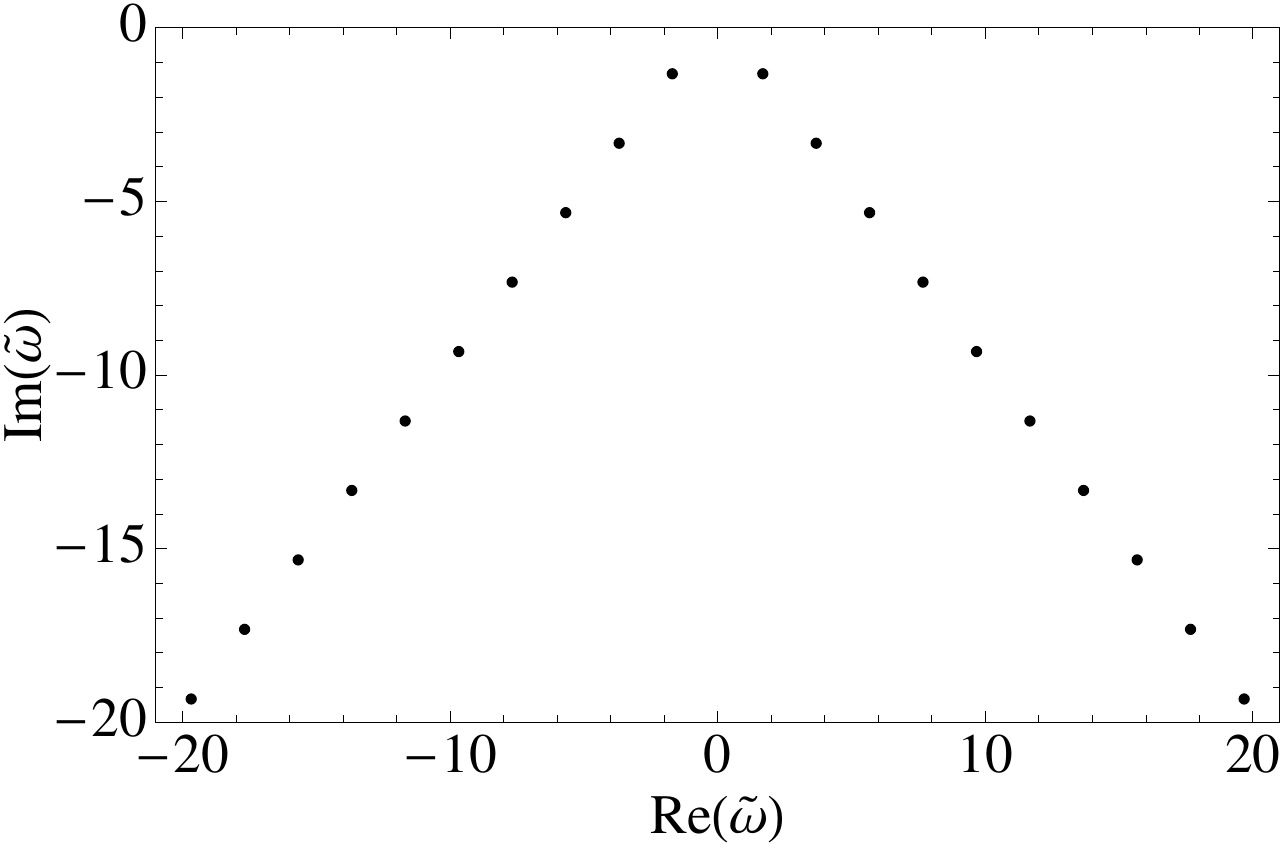}}
\end{tabular}
\caption{\label{table1}\small Quasi-normal modes for the first 11 excited states.}
\end{table}
Qualitatively the spectrum of quasi-normal modes has the same structure as the one for the D3/D7 system studied in ref. \cite{Hoyos:2006gb}. As one can see from the plot in table \ref{table1} for large $k$ the quasi-normal modes satisfy the approximate relation:
\begin{equation}
\tilde\omega_k=\pm 0.68 - .33i + (2k+1) (\pm1-i)
\end{equation}

If we apply the numerical techniques used to generate the spectrum in figure \ref{fig:fig11} to obtain the spectrum presented in table \ref{table1} we find that the results agree with relative error of $0.01\%$ (for the lowest quasi-normal modes). A more complete and detailed study of the spectrum of  quas-normal modes for non-zero magnetic field is beyond the scope of this paper. However analysis of equation (\ref{thet0}) shows that as the magnetic field increases the lowest lying modes given in table \ref{table1} shift radially away from the origin. Furthermore a new set of quasi-normal modes with vanishing real part emerges. The lowest lying of these modes is the one analyzed in the previous subsection. As one can see from figure \ref{fig:fig11} for sufficiently large magnetic fields this mode becomes tachyonic. If we keep on increasing the magnetic field more and more tachyonic modes emerge. In the strict $\eta\to\infty$ limit (corresponding to vanishing temperature) the spectrum contains an infinite tower of tachyonic modes. The qualitative behaviour is similar to the one described in \cite{Filev:2007qu} where the zero temperature spectrum for the D3/D7 set up was explored. 

\subsection{Fluctuations along $\phi$ and $\psi$}

In this subsection we study fluctuations along $\phi$ and $\psi$. Since translations along $\phi$ and $\psi$ correspond to the generators of the spontaneously broken $SU(2)$ symmetry for small bare masses we expect to detect pseudo-Goldstone modes. This is why we focus on the spectrum in the confined phase. The spectrum at zero temperature has been extensively studied in ref. \cite{Filev:2009xp}. We expect that the qualitative behaviour at finite temperature will remain the same in the confined phase.

\begin{figure}[h] 
\centering
\includegraphics[width=12cm]{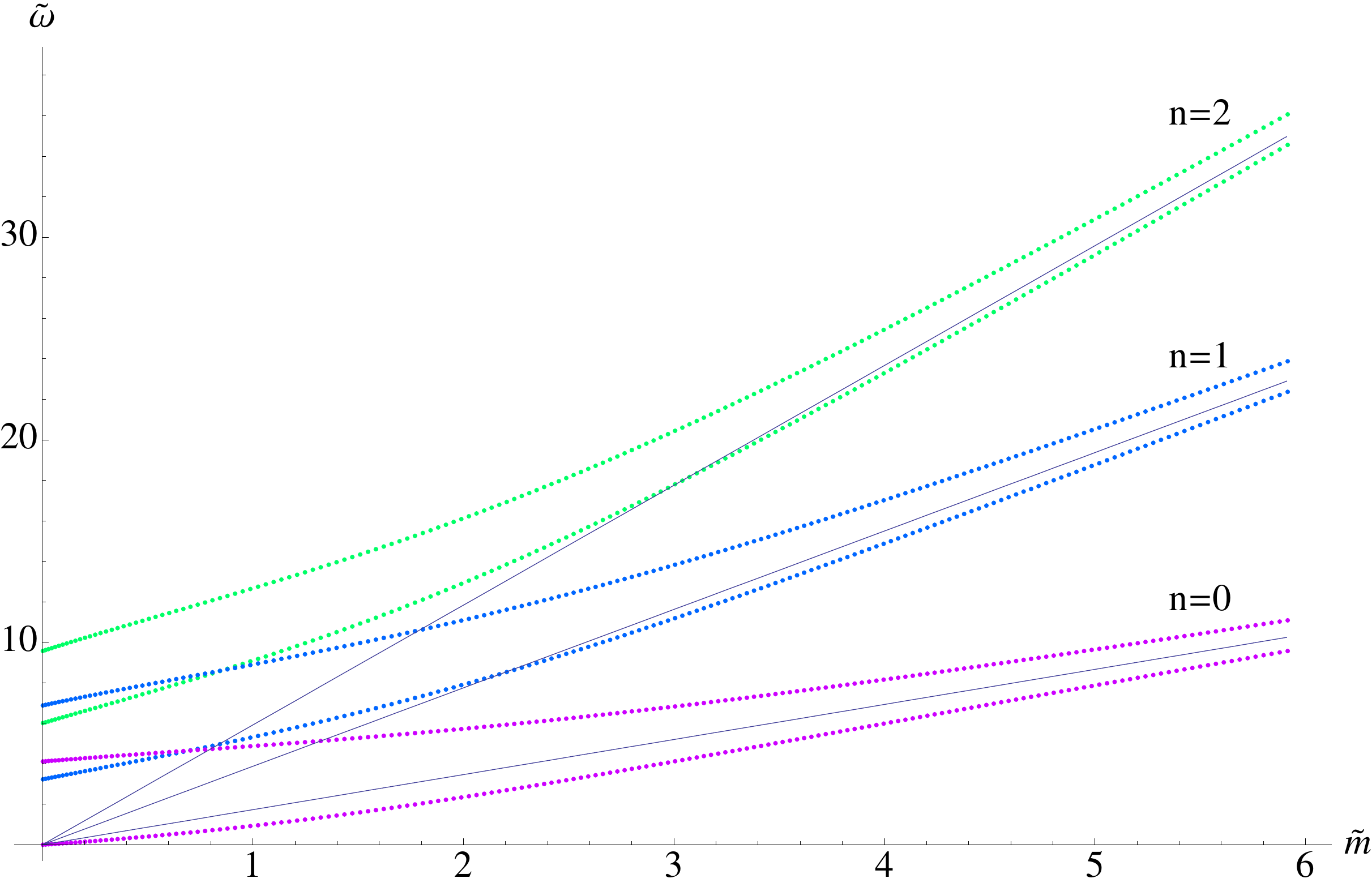}
\caption{\small A plot of $\tilde\omega$ versus $\tilde m$. One can observe a Zeeman splitting of the energy levels at large bare masses $\tilde m$. And the existence of a pseudo-Goldstone mode at small bare masses.}
\label{fig:fig12}
\end{figure}

The equations of motion derived from the quadratic action
(\ref{qvfl}) are the following:
\begin{eqnarray}
&&\partial_{\rho}\left(\frac{g(\rho)l^2}{1+l'^2}\partial_{\rho}\delta\psi\right)+\frac{4g(\rho)R^4l^2}{4r^4+b^4}\tilde\Box\delta\psi+\frac{g(\rho)l^2}{\rho^2}\Delta_{(2)}\delta\psi-PH\partial_0\delta\phi=0\ ,\\
&&\partial_{\rho}\left(\frac{g(\rho)l^2}{1+l'^2}\partial_{\rho}\delta\phi\right)+\frac{4g(\rho)R^4l^2}{4r^4+b^4}\tilde\Box\delta\phi+\frac{g(\rho)l^2}{\rho^2}\Delta_{(2)}\delta\phi+PH\partial_0\delta\psi=0\ ,\nonumber
\end{eqnarray}
where $\tilde\Box$ is give in equation (\ref{Box}). Next we consider a plane-wave ansatz:
\begin{equation}
\delta\phi=e^{i(\omega t-\vec k\dot \vec x)}\eta_1(\rho);~~~\delta\psi=e^{i(\omega t-\vec k\dot \vec x)}\eta_2(\rho)\label{anzatsqv}\, ,
\end{equation}
now using the anzatz (\ref{anzatsqv}) we get:
\begin{eqnarray}
&&\partial_{\rho}\left(\frac{g(\rho)l^2}{1+l'^2}\eta_1'\right)+\frac{4g(\rho)R^4l^2}{4r^2+b^4}\left(\frac{\omega^2}{1-\frac{16b^4r^4}{(4r^4+b^4)^2}} -\frac{\vec k^2}{1+\frac{16R^4H^2r^4}{(4r^4+b^4)^2}}\right)\eta_1-i\omega PH\eta_2=0\ ,\label{coupled}\\
&&\partial_{\rho}\left(\frac{g(\rho)l^2}{1+l'^2}\eta_2'\right)+\frac{4g(\rho)R^4l^2}{4r^4+b^4}\left(\frac{\omega^2}{1-\frac{16b^4r^4}{(4r^4+b^4)^2}} -\frac{\vec k^2}{1+\frac{16R^4H^2r^4}{(4r^4+b^4)^2}}\right)\eta_2+i\omega PH\eta_1=0\ .\nonumber
\end{eqnarray}
The equations of motion in (\ref{coupled}) can be decoupled by the
definition $\eta_{\pm}=\eta_1\pm i\eta_2$. The result is:
\begin{eqnarray}
&&\partial_{\rho}\left(\frac{g(\rho)l^2}{1+l'^2}\eta_+'\right)+\frac{4g(\rho)R^4l^2}{4r^4+b^4}\left(\frac{\omega^2}{1-\frac{16b^4r^4}{(4r^4+b^4)^2}} -\frac{\vec k^2}{1+\frac{16R^4H^2r^4}{(4r^4+b^4)^2}}\right)\eta_+-\omega PH\eta_+=0\ ,\label{decoupled}\\
&&\partial_{\rho}\left(\frac{g(\rho)l^2}{1+l'^2}\eta_-'\right)+\frac{4g(\rho)R^4l^2}{4r^4+b^4}\left(\frac{\omega^2}{1-\frac{16b^4r^4}{(4r^4+b^4)^2}} -\frac{\vec k^2}{1+\frac{16R^4H^2r^4}{(4r^4+b^4)^2}}\right)\eta_-+\omega PH\eta_-=0\ .\nonumber
\end{eqnarray}
We can now solve numerically the equations of motion by imposing boundary conditions:
\begin{equation}
\eta_{\pm}(\epsilon)=1;~~\eta_{\pm}'(\epsilon)=0;
\end{equation}
For sufficiently small $\epsilon\sim10^-6$. The spectrum is quantized by requiring that $\eta_{\pm}\sim 1/\rho$ for large $\rho$. Since we are interested in the confined phase we consider $\eta=4.5>\eta_{\rm{cr}}$. The resulting spectrum is presented in figure \ref{fig:fig12},
where we have introduced dimensionless parameters ${\tilde{m}=m/b}$ and ${\tilde{w}=R^2\omega/b}$. As one can see for large bare masses there is a Zeeman splitting of the energy levels, and the spectrum approximates to the spectrum for zero temperature and magnetic field studied in refs.~\cite{{Arean:2006pk},{Myers:2006qr}}, where the authors
obtained the following relation:
\begin{equation}
\omega_n=\frac{2m}{R^2}\sqrt{(n+1/2)(n+3/2)}\ ,\label{purespect}
\end{equation}
between the eigenvalue of the $n^{th}$ excited state $\omega_n$ and the bare mass $m$. The black fitting lines in figure \ref{fig:fig12} correspond to equation (\ref{purespect}). It is also evident that at zero bare mass the ground state has zero frequency. This is the Goldstone mode of the broken $SU(2)$ symmetry. Note that we have broken two generators while we observe only a single Goldstone mode. This is the same behaviour as in the zero temperature case studied in ref.~\cite{Filev:2009xp}. The apparent contradiction is clarified by the observation that the SO(1,2) Lorentz symmetry is broken down to $SO(2)$ rotational symmetry (by both the external magnetic field and the finite temperature). This
opens the possibility of having two types of Goldstone modes: type I
and type II satisfying odd and even dispersion relations
correspondingly. In this case there is a modified counting rule
(ref.~\cite{Nielsen:1975hm})
which states that {\it the number of GBs of type I plus twice the
  number of GBs of type II is greater than or equal to the number of
  broken generators.}

\begin{figure}[h] 
\centering
\includegraphics[width=7cm]{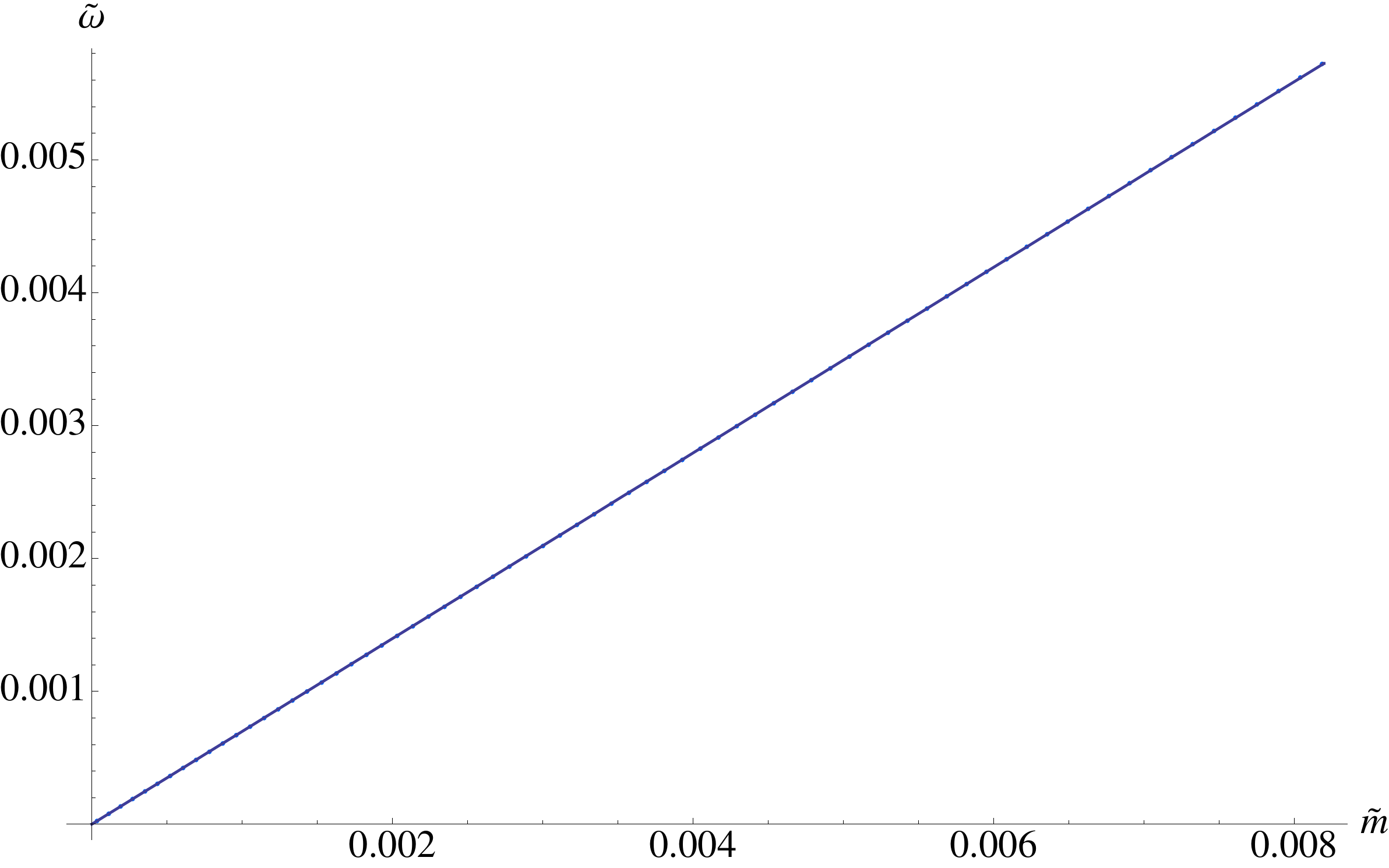}
\hspace{0.5cm}
\includegraphics[width=7cm]{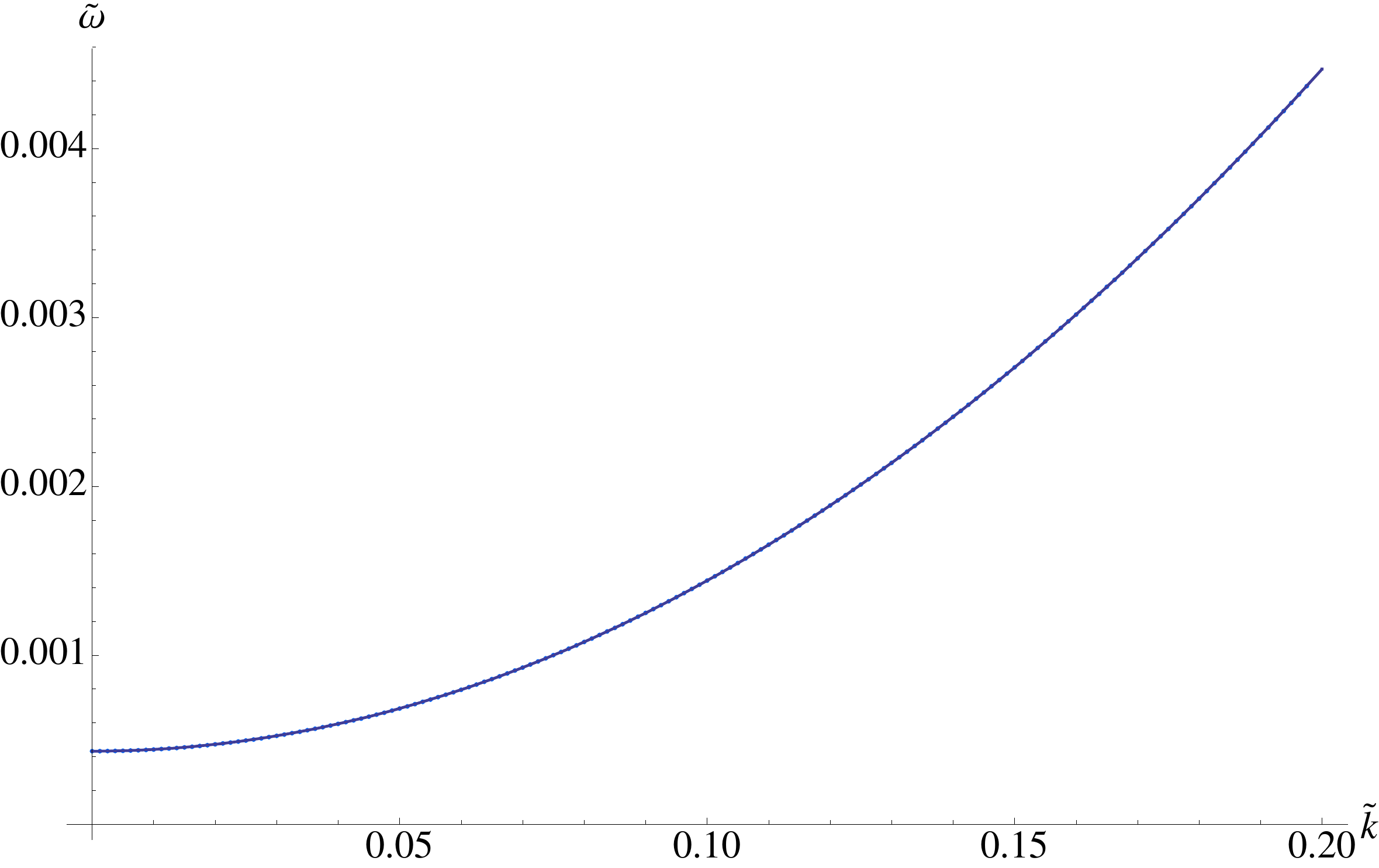}
\caption{\small Plots of the linear relation $\tilde\omega\propto\tilde m$ and of the non-relativistic dispersion relation of the pseudo-Goldstone modes.}
\label{fig:fig13}
\end{figure}

Another interesting feature of the set up is that for small bare masses the pseudo-Goldstone modes satisfy a modified Gell-Mann-Oaks-Renner relation \cite{GellMann:1968rz}. Indeed as it can be seen from the plot in figure \ref{fig:fig13} we have a linear relation (as opposed to square root one) between the frequency $\tilde\omega$ and the bare mass $\tilde m$. 
It turns out that the slope of the observed linear relation is given by:
\begin{equation}
\tilde\omega=\frac{4}{\pi\eta}\tilde c_{\rm{0}}\tilde m\ ,
\end{equation}
where $\tilde c_{\rm{0}}$ represents the condensate at zero bare mass. In fact one can prove a more general result:
\begin{equation}
\tilde\omega=\gamma{\vec{\tilde k}}^2+\frac{4}{\pi\eta}\tilde c_{\rm{0}}\tilde m\ ,\label{disprel}
\end{equation}
where
\begin{equation}
\gamma=\frac{4}{\pi\eta}\int\limits_0^{\infty}d\tilde\rho\frac{4g(\tilde\rho)\tilde l^2(4\tilde r^4+1)}{(4\tilde r^4+1)^2+16\eta^2\tilde r^4}\label{gama}
\end{equation}
and we have defined $\vec{\tilde k}=R^2 {\vec k}/b$. Let us first verify the dispersion relation from equation (\ref{disprel}) numerically. To do this end we consider a D5-brane embedding with a very small bare mass $\tilde m\approx 0.00062$. Next we generate a plot of $\tilde\omega$ versus $\tilde k$ for fixed $\eta=4.5$. The resulting plot is presented in figure \ref{fig:fig13}.
The black fitting curve corresponds to the relation:
\begin{equation}
\tilde\omega\approx0.0004326+ 0.1010 {\vec{\tilde k}}^2
\end{equation}
On the other side the expression for $\gamma$ form equation (\ref{gama}) is $\gamma\approx0.1011$ and by calculating numerically $\tilde c_0$ and for $\tilde m\approx 0.00062$ we obtain $4\tilde c_{\rm{0}}/\pi\approx0.0004321$. One can see that the relative error is $0.06\%$ and $0.05\%$ correspondingly. The analytic prove of the dispersion relation in equation (\ref{disprel}) is a generalization of the zero temperature case considered in ref.~\cite{Filev:2009xp}. Let us briefly provide the proof.
\subsubsection{Low energy dispersion relation}
 Let us consider the limit of small $\omega$ in equation (\ref{decoupled}) thus
leaving only the linear potential term in $\omega$. 
\begin{equation}
\partial_{\tilde\rho}\left(\frac{g(\tilde\rho)\tilde l^2}{1+\tilde l'^2}\eta_+'\right)-\left(\tilde\omega \tilde P\eta+\frac{4g(\tilde\rho)\tilde l^2(4\tilde r^4+1)}{(4\tilde r^4+1)^2+16\eta^2\tilde r^4}{\vec {\tilde k}}^2\right)\eta_+=0\label{eqet+}
\end{equation}
Note that we have written down equation (\ref{eqet+}) in the dimensionless variables defined previously and used the notation $\tilde P= P/R^4$. 
It is convenient to define the following variables:
\begin{equation}
\Theta^2=\frac{g(\tilde\rho)\tilde l^2}{1+\tilde l'^2}\ ;~~~\xi=\eta_+\Theta\ .
\end{equation}
Then equation~(\ref{eqet+}) can be written as:
\begin{equation}
\ddot\xi-\frac{\ddot\Theta}{\Theta}\xi-\left(\tilde\omega \tilde P\eta+\frac{4g(\tilde\rho)\tilde l^2(4\tilde r^4+1)}{(4\tilde r^4+1)^2+16\eta^2\tilde r^4}{\vec {\tilde k}}^2\right)\frac{\xi}{\Theta^2}=0\ .\label{xi}
\end{equation}
Where the overdots represent derivatives with respect to $\rho$. Now if we take
the limit $\tilde m\to 0$ and $\tilde k\to 0$ we have that $\tilde \omega\to 0$ and obtain that:
\begin{equation}
\xi=\Theta|_{\tilde\omega=0}\equiv\bar\Theta\ ,
\end{equation}
is a solution to equation (\ref{xi}). Our next step is to consider small $\tilde m$ and expand:
\begin{equation}
\xi=\bar\Theta+\delta\xi\ ;~~~\Theta=\bar\Theta+\delta\Theta\ ,\label{expcrit}
\end{equation}
where the variations $\delta\xi$ and $\delta\Theta$ are vanishing in
the $\tilde m\to 0$ limit. Then, to leading order in $\tilde m$ (keeping in mind
that $\tilde\omega\sim \tilde m$ and $\vec \tilde k^2\sim \tilde m$) we obtain:
\begin{equation}
\delta\ddot\xi-\frac{\ddot{\bar\Theta}}{\bar\Theta}\delta\xi-\delta\left(\tilde\omega \tilde P\eta+\frac{4g(\tilde\rho)\tilde l^2(4\tilde r^4+1)}{(4\tilde r^4+1)^2+16\eta^2\tilde r^4}{\vec {\tilde k}}^2\right)\frac{1}{\bar\Theta}=0\ .
\label{prib}
\end{equation}
Now we multiply equation (\ref{prib}) by $\bar\Theta$ and integrate
along $\tilde\rho$. The result is:
\begin{equation}
(\bar\Theta\delta\dot\xi-\dot{\bar\Theta}\delta\xi)\Big |_0^{\infty}-(\bar\Theta\delta{\dot\Theta}-\dot{\bar\Theta}\delta\Theta)\Big |_0^{\infty}-\tilde\omega \eta\int\limits_0^{\infty}{d\tilde\rho}\tilde P(\tilde\rho)-\frac{\pi\eta}{4}\gamma{\vec{\tilde k}}^2=0\ .\label{intermid}
\end{equation}
Using the definitions of $\Theta, \tilde P(\tilde\rho)$ and $\xi$ and requiring regularity at infinity for $\eta_+$, one can show that the first term in equation~(\ref{intermid}) vanishes and that:
\begin{equation}
 (\bar\Theta\delta{\dot\Theta}-\dot{\bar\Theta}\delta\Theta)\Big |_0^{\infty}=\tilde c\delta\tilde m\ ;~~~\int\limits_0^{\infty}{d\tilde\rho}\tilde P(\tilde\rho)=-\pi/4\ ,
\end{equation}
and hence using that $\delta\tilde m=\tilde m$ we obtain equation (\ref{disprel}) which we duplicate below:
\begin{equation}
\tilde\omega=\gamma{\vec{\tilde k}}^2+\frac{4}{\pi\eta}\tilde c_{\rm{0}}\tilde m\ ,\label{disprel2}
\end{equation}
Further analysis of the pseudo goldstone spectrum would involve derivation of the effective chiral action along the lines of ref.~\cite{Filev:2009xp}. We leave such studies for future investigations.

\section{Conclusion}

In this paper we studied flavoured holographic 1+2 dimensional Yang-Mills theory at finite temperature in an external magnetic field.
One of the main results of our studies was the construction of the phase diagram of the theory in Section 3. It seems that the observed structure of the phase diagram can be understood based on rather general grounds. At a weak magnetic field the observed quadratic behaviour in figure \ref{fig:fig4} can be understood as representing the fact that the theory is describable by a Born-Infield like action and hence the free energy is an even function of the magnetic field. The square root behaviour at large magnetic fields on the other side represents competition of energy scales and thus is consequence of the freezing effect that the magnetic field has on the phase transition. Perhaps this could explain why the phase diagram in the coordinates used in figure \ref{fig:fig3} has the same shape as the phase diagram of the 1+3 dimensional case studied in refs.~\cite{Albash:2007bk,Erdmenger:2007bn}. The similarity with the phase diagram of the Gross-Neveu model in 1+2 dimensions studied in refs.~\cite{Klimenko:1991he,Klimenko:1991he} could perhaps also be interpreted along these lines. It would be interesting to use alternative non-perturbative techniques to study the phase diagram of the defect field theory holographically dual to our set up and compare with the result obtained via the AdS/CFT correspondence.

Interesting feature of the theory is the observation that the magnetic field and the temperature have the same effect on the theory for large bare masses. This is to be contrasted to the results for the D3/D7 set up. More precisely this suggests that one can tune the sign of the condensate at large bare masses.  Furthermore, one has analytic control on that regime of the theory. It is worth looking for possible applications of this property of the system.

Another interesting feature of the theory is the enhanced relative jump of the entropy at zero bare mass. It would be interesting to employ alternative methods and explore the jump of entropy. For example by counting how the number of microstates changes. It is also intriguing that the ratio of the entropy before and after the phase transition seems to be numerically very close to $2\pi$ suggesting that perhaps this ratio can be obtained in a closed form.

A possible direction for future studies is the derivation of the temperature dependence of the condensate of the theory at zero bare mass, using alternative field theory approach. In general such studies could verify the validity of the holographic set up.

An interesting property of the D3/D5 set up is that the holographic renormalization of the D5-brane action does not require additional counter terms due to the external electric or magnetic field. This enables one to define unambiguously the magnetization of the theory. This observation suggests that D5-brane probing of more complicated asymptotically AdS$_5$ geometries would still have the same property.  It would also be interesting to further analyze the temperature dependence of the diamagnetic response in the deconfined phase and attempt to model that dependence based on the control that we have on the temperature dependence of other properties of the theory, such as the conductivity.

The meson spectrum of the theory could also be analyzed in more details. For example a more complete study of the spectrum of quasi-normal modes including non-zero bare masses. Finally, studies of the effective chiral action along the lines of the studies performed in \cite{Filev:2009xp} are of potential phenomenological interest. 

\section{Acknowledgments}

V. F. would like to thank Johanna Erdmenger, Hovhannes Grigoryan, Deog Ki Hong, Clifford Johnson, Stefano Kovacs, Arnab Kundu, Rene Meyer, Andy O'Bannon, Denjoe O'Connor, Babar Qureshi, Jonathan Shock, Plamen Stamenov and Andrew Zayakin for useful comments and suggestions. V. F. would like to thank the Aspen Center for Theoretical Physics for hospitality at the early stages of this project.
The work of V.F. was supported by an IRCSET Postdoctoral Fellowship.

\end{document}